\documentclass[AMA,sort&compress,hidelinks,LATO1COL]{WileyNJD-v2}

\articletype{Special Issue Research Article}%

\usepackage{amssymb}

\DeclareMathOperator*{\argmin}{arg\,min}

\begin{document}

\title{Multidimensional Correlation Spectroscopic Imaging of Exponential Decays: From Theoretical Principles to In Vivo Human Applications}

\author[1,2]{Daeun Kim*}

\author[3,4]{Jessica L. Wisnowski}

\author[5,6,7]{Christopher T. Nguyen}

\author[1,2]{Justin P. Haldar}

\authormark{D. Kim \textsc{et al}}

\address[1]{\orgdiv{Ming Hsieh Department of Electrical and Computer Engineering}, \orgname{University of Southern California}, \orgaddress{\state{CA}, \country{USA}}}

\address[2]{\orgdiv{Signal and Image Processing Institute}, \orgname{University of Southern California}, \orgaddress{\state{CA}, \country{USA}}}

\address[3]{\orgdiv{Radiology}, \orgname{Children's Hospital Los Angeles}, \orgaddress{\state{CA}, \country{USA}}}

\address[4]{\orgdiv{Pediatrics}, \orgname{Children's Hospital Los Angeles}, \orgaddress{\state{CA}, \country{USA}}}

\address[5]{\orgdiv{Harvard Medical School and Cardiovascular Research Center}, \orgname{Massachusetts General Hospital}, \orgaddress{\state{MA}, \country{USA}}}

\address[6]{\orgdiv{Martinos Center for Biomedical Imaging, Radiology}, \orgname{Massachusetts General Hospital}, \orgaddress{\state{MA}, \country{USA}}}

\address[7]{\orgdiv{Biomedical Imaging Research Institute}, \orgname{Cedars-Sinai Medical Center}, \orgaddress{\state{CA}, \country{USA}}}

\corres{*Daeun Kim, \email{daeunk@usc.edu}}


\abstract[Abstract]{
Multiexponential modeling of relaxation or diffusion MR signal decays is a popular approach for estimating and spatially mapping different microstructural tissue compartments.  While this approach can be quite powerful, it is also limited by the fact that one-dimensional multiexponential modeling is an ill-posed inverse problem with substantial ambiguities.  In this paper, we present an overview of a recent multidimensional correlation spectroscopic imaging approach to this problem. This approach helps to alleviate ill-posedness by leveraging multidimensional contrast encoding (e.g., 2D diffusion-relaxation encoding or 2D relaxation-relaxation encoding) combined with a regularized spatial-spectral estimation procedure.     Theoretical calculations, simulations, and experimental results are used to illustrate the benefits of this approach relative to classical methods.  In addition, we demonstrate an initial proof-of-principle application of this kind of approach to in vivo human MRI experiments.
}


\maketitle

\footnotetext{\textbf{Abbreviations:} ADMM, alternating directions method of multipliers; CPMG Carr-Purcell-Meiboom-Gill; CRLB, Cram\'{e}r-Rao lower bound; CSF, cerebrospinal fluid; GM, gray matter; ILT, inverse Laplace transform; IR-MSE, inversion-recovery multi-echo spin-echo; MRI, magnetic resonance imaging; NNLS, nonnegative least squares; PET, positron emission tomography; SNR, signal-to-noise ratio; TE, echo time; TI, inversion time; WM, white matter}

\section{Introduction}\label{sec:intro}
MRI is a powerful and versatile imaging modality, but ever since it was first introduced, its capabilities have always been practically limited by undesirable trade-offs between spatial resolution, signal-to-noise ratio (SNR), and data acquisition time.  Due to these limitations, modern human MRI experiments are typically performed with millimeter-scale voxels, even though many scientifically- or clinically-interesting biological tissue features would only become directly visible at finer (e.g., microscopic or cellular) resolution scales.  However, MRI-based study of tissue microstructure is still possible by leveraging the fact that certain MR contrast mechanisms, e.g., those based on diffusion characteristics  \cite{jones2011,tournier2011,LeBihan2012} or relaxation characteristics \cite{fenrich2001, Koenig1990, mackay2006, Deoni2010, Does2018}, are sensitive to features of the local tissue microenvironment.  This means that information about sub-voxel tissue features may still be accessible by formulating and solving an appropriate inverse problem.

Multicomponent modeling of relaxation or diffusion decay curves \cite{Kroeker1986, Whittall1989, fenrich2001, Mackay1994, Labadie1994,  Whittall1997, Does2002, Lancaster2003, mackay2006, Du2007, Deoni2008, Niendorf1996, Mulkern1999, Yablonskiy2003, Assaf2004, Assaf2008, Wang2011, Zhang2012, Scherrer2015, Kaden2016} is one of the most common approaches for resolving sub-voxel microstructure.  The basic assumption of these approaches is that a large macroscopic voxel can be modeled as containing multiple different ``compartments'' corresponding to the water pools from distinct tissue microenvironments, where each compartment is likely to exhibit distinct diffusion or relaxation decay characteristics.  As a result, neglecting any inter-compartmental exchange, the measurements from a single voxel can be modeled as a partial volume mixture of the distinct relaxation or diffusion signatures (which generally take the form of exponential decays) that would be observed from each of the sub-voxel compartments.  

While some methods have used a small preselected number of compartments based on prior assumptions about tissue characteristics \cite{Lancaster2003,Du2007,Niendorf1996,Mulkern1999}, a more flexible approach (which can accommodate an arbitrary and a priori unknown number of compartments) is to model the signal from a single voxel as a continuous distribution (or ``spectrum'') of different exponential decays \cite{Kroeker1986, Whittall1989, fenrich2001, Mackay1994, Labadie1994, Whittall1997,Does2002, Lancaster2003, Yablonskiy2003, Scherrer2015}.  In practical applications, these spectra generally exhibit distinct peaks that are usually ascribed to distinct compartments, and it is common to use spectral peak integrals to measure the contributions from each compartment.  In imaging experiments, it is common to form a ``spectroscopic image'' that consists of a distinct decay spectrum for every spatial location, and the spatial maps of the spectral peak integrals can provide important additional insights into the spatial organization of the tissue compartments \cite{Mackay1994,Labadie1994,Whittall1997}.

With 1D contrast encoding (i.e., designed to provide spectral information about a single decay parameter such as the $T_1$, $T_2$, or $T_2^*$ relaxation parameters or the apparent diffusion coefficient $D$), the inverse problem associated with estimating the exponential decay spectrum for a given voxel is sometimes described as a kind of 1D limited-data Inverse Laplace Transform (ILT)\cite{istratov1999}.  Unfortunately, it has been recognized for centuries that this inverse problem is fundamentally ill-posed and difficult to solve \cite{istratov1999,prony1795}.  Practically, this ill-posedness means that it is extremely difficult to separate tissue compartments that have similar exponential decay characteristics.

To avoid some of the ill-posedeness associated with 1D spectrum estimation, multidimensional contrast encoding has been investigated in many applications \cite{Peemoeller1981, English1991,  Saab2001, Hurlimann2002, Song2002, Does2002, Callaghan2003, galvosas2010, freed2010, Bernin2013, Celik2013, Bai2015, Benjamini2016, Slator2019}.  These approaches acquire a high-dimensional dataset that nonseparably encodes two or more decay parameters (e.g., joint $D$-$T_2$ contrast encoding \cite{Hurlimann2002, Callaghan2003} or joint $T_1$-$T_2$ contrast encoding \cite{Peemoeller1981, English1991}), and then solve an inverse problem to estimate a multidimensional correlation spectrum that describes the joint distribution of these multiple decay parameters within each voxel. This work has successfully demonstrated that multidimensional encoding and multidimensional spectrum estimation has benefits over 1D approaches.

The early work on multidimensional correlation spectroscopy of exponential decays has usually reported correlation spectra representing large spatial volumes.  These were obtained by either exciting and collecting signal from a large spatial volume at once with no additional spatial encoding, or by acquiring imaging data and then averaging the measured signal over regions of interest.  Notably, there were no attempts to perform spatial mapping of the integrated spectral peaks until very recently.  In principle, it would have been straightforward to acquire imaging data and generate spatial maps by performing voxel-by-voxel estimation of the multidimensional correlation spectra.  However in practice, while the multidimensional inverse problem is not as ill-posed as the 1D problem, it is still somewhat ill-posed.  As a consequence, conventional spectrum estimation approaches are associated with very onerous data quantity and quality requirements that would be hard to satisfy in a relatively short-duration imaging experiment with relatively good spatial resolution.  

However, advances in multidimensional data sampling design and multidimensional correlation spectrum estimation techniques have recently reduced data quantity and quality requirements, which has enabled some of the first reports of spatial mapping based on the spectral integrals from this type of multidimensional experiment  \cite{Kim2016isrmrm,Kim2017, Kim2017assilomar, Kim2018isbi, Benjamini2017, Slator2019}.

In this paper, we will present an overview of an approach that we developed to enable spatial mapping of the spectral peaks from multidimensional correlation spectra \cite{Kim2016isrmrm,Kim2017, Kim2017assilomar, Kim2018isbi}.  This approach relies on the use of spatial-spectral regularization and estimation-theoretic multidimensional sampling design. Recent estimation approaches developed by other groups are also promising, and are potentially complementary to our approach because they leverage different assumptions about the multidimensional spectrum to enable spatial mapping \cite{Benjamini2017}. 

The approach that we developed has recently enabled spatial maps derived from multidimensional $D$-$T_2$ spectra in physical phantoms and ex vivo tissue samples \cite{Kim2016isrmrm,Kim2017}.  For the present paper, we demonstrate spatial maps derived from multidimensional $T_1$-$T_2$ spectra, including what we believe to be the first in vivo human brain results from this kind of experiment.  Some preliminary accounts of our $T_1$-$T_2$ experiments were previously presented in recent conferences \cite{Kim2017spie,Kim2018ismrm, Kim2018isbi, Kim2017assilomar}.

\section{Theory and Methods}

\subsection{Exponential Decay Spectroscopy and the Benefits of Multidimensional Spectral Information}\label{sec:spect}

As noted in the introduction, many methods choose to represent the measured signal from a macroscopic voxel in an experiment with 1D contrast encoding as a spectrum of exponential decays, where the peaks of the spectrum are ascribed to different compartments.  For the sake of concreteness, our description in the remainder of this paper will present descriptions corresponding to $T_1$ and $T_2$ relaxation parameters, although similar modeling principles also apply to other important MR decay parameters such as $T_2^*$ and $D$.  

In 1D $T_2$-relaxometry based on a spin-echo acquisition \cite{Kroeker1986, Whittall1989},  the ideal noiseless signal from a large voxel can be modeled as
\begin{equation}
m(TE) = \int f(T_2) e^{-TE/T_2} dT_2,\label{eq:t2}
\end{equation}
where $m(TE)$ is the ideal observed signal as a function of echo time (TE), and $f(T_2)$ is the continuous 1D spectrum of $T_2$ relaxation parameters that needs to be estimated from the data.   Similarly, in $T_1$-relaxometry based on an inversion recovery sequence \cite{Kroeker1986, Labadie1994}, the ideal noiseless signal from a large voxel can be modeled as
\begin{equation}
m(TI) = \int f(T_1) (1-2e^{-TI/T_1}) dT_1,\label{eq:t1}
\end{equation}
where $m(TI)$ is the ideal observed signal as a function of inversion time (TI),  and $f(T_1)$ is the continuous 1D spectrum of $T_1$ relaxation parameters that needs to be estimated from the data.  In both cases, the 1D spectrum model reflects the implicit assumption that a macroscopic imaging voxel may be viewed as the linear mixture of a large (potentially infinite!) number of  microenvironments with distinct exponential decay characteristics and negligible inter-compartmental exchange.  And in both cases, data is acquired by varying a 1D contrast-encoding parameter, e.g., either TE or TI.  Estimating the 1D spectrum from either Eq.~\eqref{eq:t2} or \eqref{eq:t1} can be described as a form of 1D limited-data ILT, which is classically ill-posed as noted previously.  Because of this ill-posedness, it is common to use additional constraints when solving these inverse problems to help stabilize the solution.  It is especially common to assume that the spectra $f(T_1)$ and $f(T_2)$ should be nonnegative, which leads to a nonnegative least squares (NNLS) \cite{Lawson1995} formulation of the inverse problem \cite{Whittall1989}.

Multidimensional correlation spectroscopy methods are based on the synergistic higher-dimensional combination of these kinds of lower dimensional models.  For example, the signal for 2D $T_1$-$T_2$ correlation spectroscopy using an inversion-recovery spin-echo pulse sequence \cite{Song2002} can be modeled as
\begin{equation}
m(TE,TI) = \int \int f(T_1,T_2)(1-2e^{-TI/T_1})e^{-TE/T_2}~dT_1dT_2,
\end{equation}
where it now becomes necessary to acquire 2D contrast-encoded data $m(TE,TI)$ in order to estimate the 2D $T_1$-$T_2$ correlation spectrum $f(T_1,T_2)$.  This 2D approach is easily generalized to even higher dimensions, e.g., by using 3D contrast encoding with a 3D $T_1$-$T_2$-$D$ correlation spectrum or by using 4D contrast encoding with a 4D $T_1$-$T_2$-$D$-$T_2^*$ correlation spectrum.   In all of these settings, classical multidimenstional spectrum estimation approaches still frequently rely on nonnegativity assumptions and NNLS fitting, just like for the 1D case. 

\begin{figure}[t]
	\centering
	{\includegraphics[width=1\linewidth]{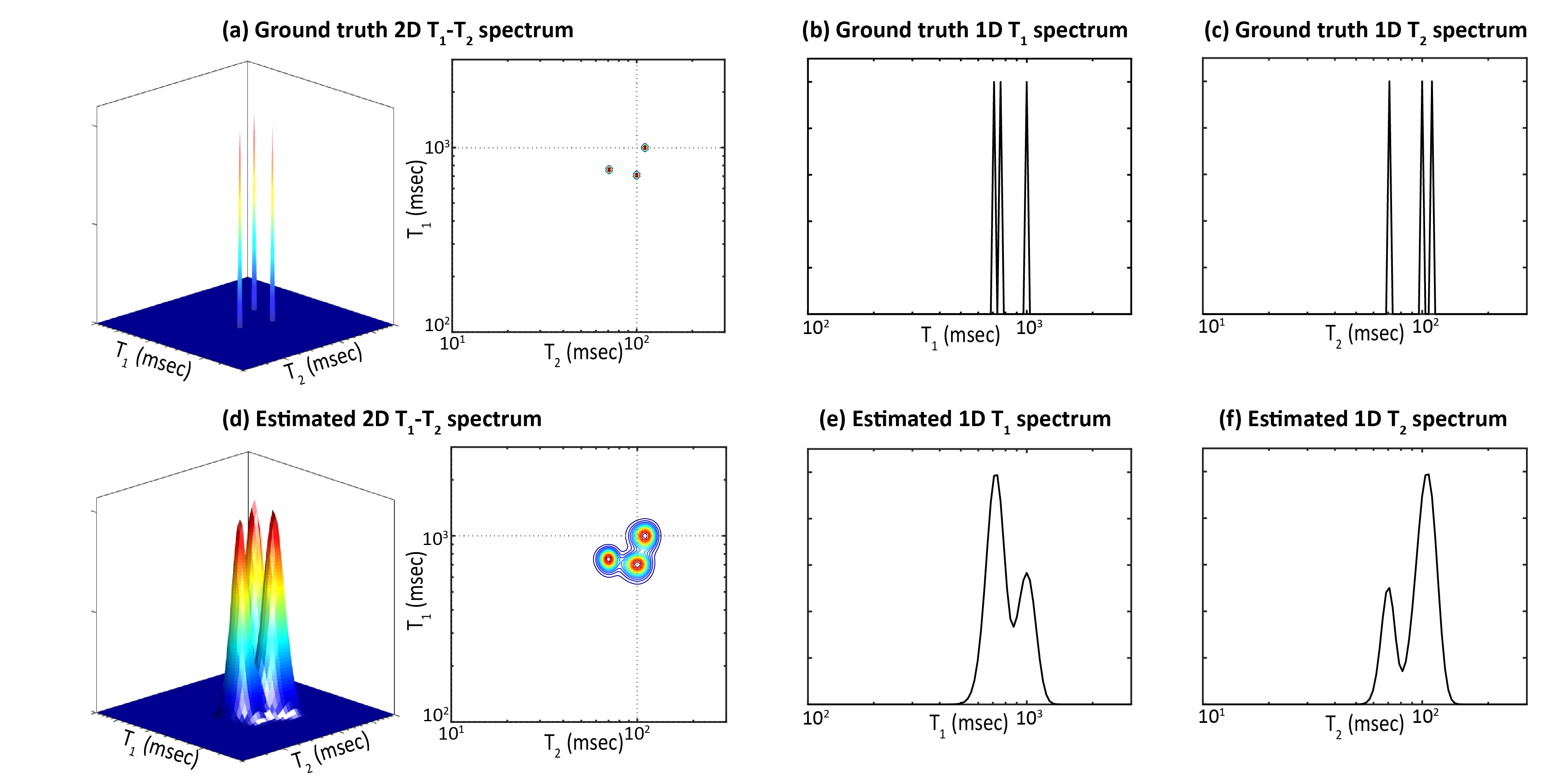}}

	\caption{Toy illustration of the advantage of 2D multidimensional correlation spectroscopy over 1D spectroscopy. Ground truth values of three spectral peaks are shown in (a) a 2D $T_1$-$T_2$ spectrum (left: a 3D plot and right: a 2D contour plot), (b) the corresponding 1D $T_1$ spectrum, and (c) the corresponding 1D $T_2$ spectrum.  While the three peaks can all be successfully resolved in these ground truth spectra, real experiments will experience degraded spectral resolution because of finite sampling and noise.  When resolution is degraded, the three peaks can still be easily resolved in (d) the 2D spectrum, though are no longer well-resolved in (e,f) either of the 1D spectra.} \label{fig:ill} 		
\end{figure}

An important advantage of multidimensional contrast encoding relative to 1D contrast encoding is reduced ill-posedness of the inverse problem.  This leads to an improved capability to successfully resolve multiple compartments, even if some of the compartments possess similar decay parameters.  For illustration, consider the toy scenario depicted in Fig.~\ref{fig:ill} in which a voxel consists of three distinct compartments, where compartment 1 has $T_2=70$ ms and $T_1 = 750$ ms, compartment 2 has $T_2 = 100$ ms and $T_1 = 700$ ms, and compartment 3 has $T_2 = 110$  ms and $T_1 = 1000$ ms.  These three spectral peaks are well-separated in the 2D $T_1$-$T_2$ space, and therefore may be relatively straightforward to resolve.  On the other hand, they are not well-separated when only considering the 1D $T_2$ dimension (i.e., compartments 2 and 3 may be hard to resolve because they have similar $T_2$ values) or only considering the 1D $T_1$ dimension (i.e., compartments 1 and 2 may be hard to resolve because they have similar $T_1$ values).

\subsubsection{Estimation Theoretic Analysis of Encoding in  Higher Dimensions}

While previous literature has confirmed the advantage of multidimensional correlation spectroscopy in this setting empirically \cite{Peemoeller1981, English1991, Saab2001, Song2002, Does2002, galvosas2010, Celik2013, Bai2015, Benjamini2016, Kim2017, Benjamini2017}, we find it instructive to examine the difference between 1D and multidimensional approaches from an estimation theoretic perspective.  Our theoretical characterization is based on the Cram\'{e}r-Rao Lower bound (CRLB), which is a theoretical lower bound on the variance of an unbiased estimator for an unknown parameter of interest  \cite{Kay1993}, and is frequently used to compare and optimize different experiment protocols in a variety of quantitative MR applications \cite{jones1996,Cavassila2001,ogg2004,pineda2005,Alexander2008,zhao2019,teixeira2018}. Since the CRLB depends on the specific values of the model parameters, we perform an illustrative  analysis  for the  case of estimating the same toy three-compartment model described above.  In this case, the ideal noiseless data for a given set of encoding parameters $(TE,TI)$ is given by
\begin{equation}
m(TE,TI) =\sum_{s=1}^{3} f_s\left( 1-2e^{-TI/T_1^s} \right) e^{-TE/T_2^s},\label{eq:sim} 
\end{equation}
where the spin density parameters $f_s$ were all set equal to 1,  and the $T_1^s$ and $T_2^s$ parameters were set to the $T_1$ and $T_2$ parameters given above for the three compartment model. Assuming white Gaussian noise, the CRLB is obtained by first computing the Fisher information matrix for the unknown parameters, and then computing the inverse of  the Fisher information matrix  \cite{Kay1993}.

When computing the CRLB for 2D $T_1$-$T_2$ correlation spectroscopy, we assumed that the number of compartments was known a priori, such that there were 9 unknown parameters to be estimated (i.e., the $f_s$, $T_1^s$, and $T_2^s$ parameters for each compartment).  For 1D $T_1$ or $T_2$ spectroscopy, we instead computed CRLBs assuming that there were 6 unknown parameters to be estimated (e.g., only the $f_s$ and $T_1^s$ need to be estimated for each compartment in the $T_1$-relaxometry case).  The 2D $T_1$-$T_2$ correlation spectroscopy acquisition was assumed to use an inversion-recovery preparation for $T_1$ contrast encoding and a Carr-Purcell-Meiboom-Gill (CPMG)  multi-spin echo sequence for $T_2$ contrast encoding, with data sampled at every combination of 7 inversion times ($TI$ = 0, 100, 200, 400, 700, 1000 and 2000) and 15 echo times ($TE$s ranging from 7.5 ms to 217.5 ms in 15 ms increments) for a total of 7 $\times$ 15 = 105 contrast encodings.  We compared against a conventional inversion recovery spin-echo sequence for $T_1$ relaxometry, using the same 7 inversion times used for the 2D case.  In this case, the expected scan time would be the same as for 2D case when the both sequences use the same TR.  We also compared against conventional $T_2$ relaxometry using a standard CPMG-based multi-spin echo sequence with 32 echo times ($TE$s ranging from 10 ms to 320 ms).  We assumed that this data was averaged 7 times, so that the experiment duration will match the duration of both the 2D correlation spectroscopy and $T_1$ relaxometry experiments. For all three experiments, the noise standard deviation was assumed to be the same.

Comparing 2D $T_1$-$T_2$ correlation spectroscopy against 1D $T_2$ relaxometry, the CRLB calculation indicates that the lower bound on the standard deviation (i.e., the square root of the CRLB) achieved for $T_2$ with 2D correlation spectroscopy  was 3.77$\times10^{2}$ times smaller for the first compartment, $1.08\times10^3$ times smaller for the second compartment, and $2.29 \times 10^3$ times smaller for the third compartment.    Comparing 2D $T_1$-$T_2$ correlation spectroscopy against 1D $T_1$ relaxometry, the CRLB calculation indicates that the lower bound on the standard deviation achieved for $T_1$ with 2D correlation spectroscopy  was 9.11$\times10^{4}$ times smaller for the first compartment, $2.21\times10^4$ times smaller for the second compartment, and $2.10 \times 10^3$ times smaller for the third compartment.  As can been seen, the CRLB values for multidimensional correlation spectroscopy are orders-of-magnitude lower than either of the conventional 1D methods, and imply that the 1D experiments would require from millions (for $T_2$) to billions (for $T_1$) times more data averaging to achieve the same CRLBs as the 2D experiment.  This calculation helps to quantify the substantial estimation theoretic improvements offered by multidimensional encoding  relative to 1D encoding.  

\subsection{Spatial-Spectral Modeling}\label{sec:spaspect}
The previous section demonstrated considerable advantages for multidimensional encoding over 1D encoding, and this advantage can be sufficient for experiments that are designed to generate spectra from large spatial volumes.  However, conventional multidimensional correlation spectroscopic imaging experiments would still have relatively onerous data quantity and quality requirements if spectral reconstruction is performed voxel-by-voxel with high spatial resolution.  One of the observations from our previous work \cite{Kim2016isrmrm,Kim2017,Kim2017spie,Kim2018ismrm, Kim2018isbi} has been that substantial gains can be achieved using spatial-spectral estimation instead of voxel-by-voxel spectral estimation.  We review these concepts below.

Without loss of generality, we consider the spatial-spectral model for 2D $T_1$-$T_2$ correlation spectroscopic imaging given by
\begin{equation}
m(\mathbf{r}, TE, TI) = \int \int f(\mathbf{r},T_1, T_2)(1-2e^{-TI/T_1})e^{-TE/T_2}~dT_1dT_2, \label{eq:spatial_spectral_model}
\end{equation}
where $m(\mathbf{r},TE,TI)$ represents the image acquired with contrast encoding parameters $(TE,TI)$ as a function of the spatial location $\mathbf{r}$, and $f(\mathbf{r},T_1, T_2)$ represents the high-dimensional spectroscopic image that is comprised of a full 2D $T_1$-$T_2$ relaxation correlation spectrum at every spatial location.  In the case where 2D imaging experiments are performed (i.e., $\mathbf{r} = (x,y)$), the spectroscopic image $f(\mathbf{r},T_1, T_2)$ would be 4D, while the spectroscopic image would be 5D for a 3D imaging experiment.  

One potential benefit of spatial-spectral modeling is that, if the spectral peak locations (but not necessarily the spin density associated with each spectral peak) are assumed to be similar for neighboring voxels, then using the data from multiple voxels to estimate the location of the shared spectral peak can substantially reduce the ill-posedness of the estimation problem.  In particular, analysis of a related problem has shown that joint spatial-spectral modeling (i.e., with the spectra for all voxels  estimated simultaneously) has the potential to reduce the the CRLB by an order of magnitude or more relative to uncoupled voxel-by-voxel estimation \cite{Lin2014}.  The benefits of spatial-spectral modeling have also been observed empirically in PET parameter estimation \cite{Lin2014} and 1D relaxometry \cite{Kumar2012,Labadie2013}, as well as our previous 2D correlation spectroscopy work \cite{Kim2016isrmrm,Kim2017,Kim2017spie,Kim2018ismrm, Kim2018isbi}.

To illustrate this improvement from an estimation theoretic perspective, it is instructive to again revisit the three compartment toy example from Section~\ref{sec:spect}.  However, instead of considering a single voxel in isolation, we now consider a scenario involving the simultaneous estimation of 2D correlation spectra from three different voxels.  For this particular toy example, we will assume that each of these three voxels has the same three compartments as in Section~\ref{sec:spect} with the same $T_1$ and $T_2$ parameter values.  However, we will assume that the volume fractions for each compartment are distinct (i.e., $f_1=f_2=f_3=1$ for voxel 1, $f_1=0.8$, $f_2 = 0.6$ and $f_3=1.8$ for voxel 2, and $f_1=2$, $f_2=0.5$, and $f_3=0.5$ for voxel 3).  Using the CRLB and the same 2D experimental paradigm from Section~\ref{sec:spect}, we compare the voxel-by-voxel estimation strategy (where the relaxation parameters are not assumed to be the same for different voxels) against a strategy that is aware that the voxels share the same relaxation parameters.  Our CRLB analysis shows that the spatial-spectral approach has CRLBs that range from 48$\times$ to 220$\times$ lower (depending on the specific parameter) relative to the voxel-by-voxel case.  The voxel-by-voxel approach would thus require hundreds of additional averages to match the good estimation theoretic characteristics of the spatial-spectral approach, again indicating a substantial reduction in the ill-posedness of the spectral estimation problem.  

While this toy example is exaggerated (since it depends on the unrealistic and very strong prior information that the compartments in all voxels have identical relaxation characteristics), it is still indicative of the potential benefits of spatial-spectral modeling.  Importantly, this additional reduction in ill-posedness implies that data quality and quantity requirements can be relaxed, which can enable high quality correlation spectroscopic imaging experiments from much shorter experiments with fewer averages and fewer contrast encodings.

\subsection{Practical Spatial-Spectral Estimation}

Based on the theory presented in the previous section, there are clear potential advantages to estimating a high-dimensional spectroscopic image from high-dimensional data using an estimation procedure that incorporates some form of spatial constraints.  However, we generally don't want to make the same very strong assumptions that were used in previous toy examples.  In the following, we describe the regularization based approach that we have developed, which is a minor variation of the approach described in our previous work\cite{Kim2016isrmrm,Kim2017,Kim2017spie,Kim2018ismrm, Kim2018isbi}.  

While the spatial-spectral model in Eq.~\eqref{eq:spatial_spectral_model} is continuous, we will use a dictionary-based discrete model for practical numerical implementation, as in conventional 1D and 2D diffusion and relaxation spectroscopy methods \cite{Kroeker1986, Whittall1989, fenrich2001, Mackay1994, Labadie1994, Whittall1997, Peemoeller1981, English1991, Saab2001, Song2002, Does2002, galvosas2010, Celik2013, Bai2015, Benjamini2016} are replaced by standard Riemann sum approximations of the form:
\begin{equation}
m(\mathbf{r}_i,TE_p, TI_p) 
= \sum_{q=1}^Q w_q f(\mathbf{r}_i,T_1^q,T_2^q)(1-2e^{-TI_p/T_1^q})e^{-TE_p/T_2^q}, \label{eq:RRCSImodel_discrete}
\end{equation}
for $\forall i=1, \ldots, N$ and $\forall p = 1, \ldots, P$.  In this expression, $N$ is the number of voxels in the image; it is assumed that we acquire $P$ different contrast encodings $(TE_p,TI_p)$; we have modeled the relaxation distribution using a dictionary with $Q$ elements, where the $q$th element corresponds to the relaxation parameters $(T_1^q, T_2^q)$; and $w_q$ is the density normalization term (i.e., the numerical quadrature weights) required for accurate approximation of the continuous integral using a finite discrete sum.  This discrete model can be equivalently represented in matrix form as
\begin{equation}
\mathbf{m}_i=\mathbf{K}\mathbf{f}_i, \label{eq:matrix_form}
\end{equation}
for $i=1,\ldots, N$, where the vector $\mathbf{m}_i\in\mathbb{R}^P$ contains all the  measured  contrast-encoded data corresponding to the $i$th voxel and has $p$th entry $[\mathbf{m}_i]_{p}=m(\mathbf{r}_i,TE_p,TI_p) $; the dictionary matrix $\mathbf{K}\in\mathbb{R}^{P \times Q}$  has entries $[\mathbf{K}]_{pq} = w_q(1-2e^{-TI_p/T_1^q})e^{-TE_p/T_2^q}$; and $\mathbf{f}_i\in\mathbb{R}^{Q}$ is the 2D spectrum corresponding to the $i$th voxel of the high-dimensional spectroscopic image and  has $q$th entry $[\mathbf{f}_i]_{q} = f(\mathbf{r}_i,T_1^q,T_2^q)$. 

Given this discrete model, we perform estimation using the same nonnegativity constraints from classical relaxation spectroscopy methods, while also using a spatial smoothness constraint on the reconstructed 2D spectra \cite{Kim2016isrmrm,Kim2017,Kim2017spie,Kim2018ismrm, Kim2018isbi}:
\begin{equation}
\left\{\hat{\mathbf{f}}_1,\hat{\mathbf{f}}_2,\ldots,\hat{\mathbf{f}}_N\right\}
= \argmin_{\{\mathbf{f}_i\in \mathbb{R}^Q\}_{i=1}^N}
\left[ \sum_{i=1}^N t_i \left \|   \mathbf{m}_i - \mathbf{K} \mathbf{f}_i \right \|_{2}^{2} + \lambda \sum_{l \in \Delta i} \left \| \mathbf{f}_i - \mathbf{f}_l \right \|_2^2 ~\right] 
\label{eq:optimization}
\end{equation}
subject to $[\mathbf{f}_i]_q \geq 0 \text{ for } \forall q=1,\ldots,Q$ and $\forall i = 1,\ldots,N$.  The first term in this expression is a standard data consistency term for each voxel.  The second term is a spatial regularization term that encourages the 2D correlation spectrum from one voxel to be similar to the 2D correlation spectra from neighboring voxels, where  $\Delta i$ is the index set for the voxels that are directly adjacent to the $i$th voxel.  The parameter $\lambda$ is a user-selected regularization parameter that controls the strength of the spatial regularization.  In the data consistency term, the variables $t_i$ correspond to a spatial mask for the object, and are equal to 0 if the $i$th voxel is outside the object and are otherwise equal to one. This spatial mask is used to avoid fitting spectra to noise-only voxels of the image, and prevents spectra within the object from being contaminated by noise when spatial smoothness constraints are imposed.  

Smoothness-based spatial regularization, as used in Eq.~\eqref{eq:optimization}, is a classical constraint that is used in a wide range of imaging inverse problems \cite{bertero1998}.  This constraint is based on the principle that the spectra and spatial maps are likely to be spatially smooth, and can be viewed as a ``soft'' way of imposing the spatial-spectral constraints described in Sec.~\ref{sec:spaspect}.  In particular, the constraint encourages spectral similarity between adjacent voxels without forcing exact correspondence, which accommodates situations in which the spectral peak locations or lineshape characteristics vary gradually from voxel to voxel.  In addition, this approach is not expected to fail in problematic ways if the spectrum from one voxel is very different from its neighbors (e.g., as will happen frequently along compartmental boundaries in the examples we show in the next section).  In such cases, the use of spatial regularization is expected to behave gracefully, e.g., by blurring a feature that may originally have been sharper \cite{bertero1998,haldar2011a}.  Importantly, the regularization parameter $\lambda$ can be varied to achieve a good balance between the ill-posedness of the estimation problem and the loss of spatial resolution, and there exist theoretical tools that can be used to quantify the trade-off between the two \cite{fessler1996,ahn2008}. 

If we let $\mathbf{F}\in \mathbb{R}^{Q \times N}$ and $\mathbf{M}\in \mathbb{R}^{P \times N}$ represent the matrices whose $i$th columns respectively correspond to $\mathbf{f}_i$ and $\mathbf{m}_i$, the optimization problem from Eq.~\eqref{eq:optimization} can also be more compactly be represented as
\begin{equation}
\hat{\mathbf{F}} = \argmin_{\mathbf{F} \in \mathbb{R}^{Q \times N}} \|\mathbf{M}\mathbf{T} - \mathbf{K}\mathbf{F}\mathbf{T}\|_F^2 + \lambda\|\mathbf{F} \mathbf{C}^H\|_F^2,\label{eq:mat}
\end{equation}
subject to $[\mathbf{F}]_{pq} \geq 0 \text{ for } \forall p=1,\ldots,P \text{ and } \forall q=1,\ldots,Q$. In this expression, $\|\cdot\|_F$ denotes the standard Frobenius norm, $\mathbf{T} \in \mathbb{R}^{N \times N}$ is a diagonal matrix whose $i$th diagonal entry contains the value of $t_i$, $\mathbf{C}$ is the matrix operator that computes spatial finite differences, and $(\cdot)^H$ denotes the Hermitian operator (conjugate transpose).  This representation is convenient for numerical optimization.

The optimization problem in  Eq.~\eqref{eq:mat}  is convex, and there are many convex optimization methods that will find the global solution from arbitrary initializations.  Our work has generally used an  alternating directions method of multipliers (ADMM) algorithm \cite{afonso2011} to solve the nonnegativity-constrained optimization problem \cite{Kim2016isrmrm,Kim2017,Kim2017spie,Kim2018ismrm, Kim2018isbi}.  We describe the steps of this algorithm in the Appendix, but refer readers to Refs.~\cite{Kim2017,afonso2011} for further detail.

\section{Examples}
As illustrative examples of multidimensional correlation spectroscopic imaging, we will demonstrate 2D $T_1$-$T_2$ correlation spectroscopic imaging using numerical simulations, real experiments with a pumpkin, and several real experiments with in vivo human brains.  While we have previously published 2D $D$-$T_2$ correlation spectroscopic imaging results \cite{Kim2017}, this is the first journal publication of 2D $T_1$-$T_2$ correlation spectroscopic imaging, as well as the first publication of in vivo multidimensional relaxation correlation spectroscopic imaging in humans.  (While preliminary accounts of some of our 2D $T_1$-$T_2$ correlation spectroscopic imaging experiments have appeared in recent conference presentations\cite{Kim2017spie,Kim2018ismrm, Kim2018isbi, Kim2017assilomar}, the present article shows substantially more datasets with more detailed analysis). 

While advanced experimental protocols would likely enable improved experimental efficiency, we have focused on a simple proof-of-principle implementation in which the 2D $T_1$-$T_2$ data is acquired with a basic inversion-recovery multi-echo spin-echo (IR-MSE) pulse sequence and 2D spatial encoding, which we model using  Eq.~\eqref{eq:spatial_spectral_model}. Subsequently, a 4D spectroscopic image is estimated by solving the optimization problem in Eq.~\eqref{eq:optimization}. 

\subsection{Numerical Simulation}
\begin{figure}[tp]
	\centering
	{\includegraphics[width=0.9\linewidth]{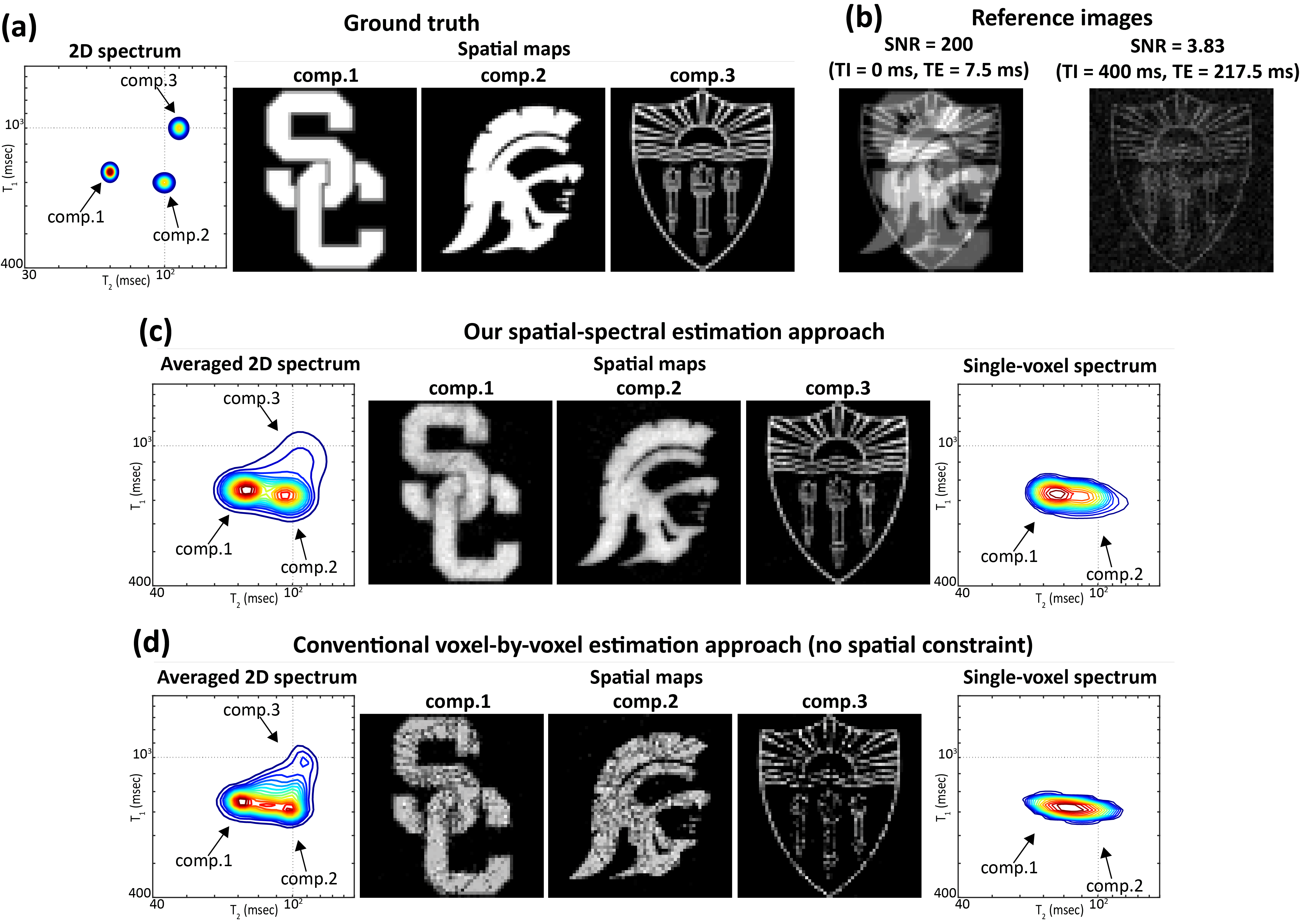}}
	
	\caption{ Numerical simulation results for 2D correlation spectrum estimation.  (a) Ground truth used for numerical simulation: (left) compartmental spectra $f_c(T_1,T_2)$ and (right) compartment spatial maps $a_c(x,y)$. (b) Representative simulated images. The highest SNR image (corresponding to $TI$ = 0 ms and $TE$ = 7.5 ms) and the lowest SNR image (corresponding to $TI$ = 400 ms and $TE$ = 217.5 ms) are displayed.  Estimation results are shown for (c) our spatial-spectral estimation approach and (d) the conventional voxel-by-voxel estimation approach (no spatial constraintd).  Each subfigure shows  (left) the 2D spectrum obtained by spatially-averaging the 4D spectroscopic image, (middle) spatial maps obtained by spectrally-integrating the spectral peak locations, and (right) a 2D spectrum obtained from a single representative voxel which contains two compartments. } \label{fig:simulation} 		
\end{figure}

Numerical simulations are valuable for understanding the characteristics of multidimensional correlation spectroscopic image estimation, because (unlike for real experiments), we have a ground truth we can compare the estimation results against.  In this simulation, a gold standard spectroscopic image was constructed using a three-compartment model according to
\begin{equation}
f(x,y,T_1,T_2) = \sum_{c=1}^{3}a_c(x,y) f_c(T_1,T_2),
\end{equation}
where $a_c(x,y)$ is the spatial distribution and $f_c(T_1,T_2)$ is the spectrum for the $c$th compartment, as shown in Figure~\ref{fig:simulation}(a).  The spectra were generated using a 2D Gaussian spectral lineshape with different spectral peak locations (compartment 1: $(T_1,T_2)$=(70ms, 750ms); compartment 2: $(T_1,T_2)$=(100ms, 700ms); compartment 3: $(T_1,T_2)$=(110ms, 1000ms)).  Note that the three compartments are difficult to distinguish using 1D $T_1$ relaxation or $T_2$ relaxation approaches because the three compartments have similar $T_1$ or $T_2$ values to one another.  Noiseless data was generated using  every combination of 7 inversion times ($TI$ = 0, 100, 200, 400, 700, 1000 and 2000 ms) and 15 echo times ($TE$s ranging from 7.5 ms to 217.5 ms in 15 ms increments) for a total of $P$ = 105 contrast encodings.  Subsequently, Gaussian noise was added to the noiseless data, and magnitudes were taken leading to Rician-distributed data.  In the dataset, image SNRs range from 3.83 to 200 (SNRs were computed separately for each contrast-encoded image as the ratio between the average per-pixel signal intensity within the image support and the noise standard deviation).  Figure~\ref{fig:simulation}(b) shows representative images from the highest SNR image (SNR = 200) with $TI$ = 0 and $TE$ = 7.5 ms and the lowest SNR image (SNR = 3.83) with $TI$ = 400 ms and $TE$ = 217.5 ms. 

For spectroscopic image estimation, a dictionary matrix $\mathbf{K}$ was formed with every combination of 100 $T_1$ values (ranging from 100 ms to 3000 ms spaced logarithmically) and 100 $T_2$ values (ranging from 2 ms to 300 ms spaced logarithmically) for a total of $Q$ = 10,000 dictionary elements.  This type of dictionary construction is typical of previous relaxation spectroscopy methods.  (Note that we also tried other dictionary constructions with different ranges for $T_1$ and $T_2$, and with $Q$ values ranging from 10,000 to 40,000. However, we only report results from this single dictionary for simplicity, because the results did not change in consequential ways when other dictionaries were used.)  Optimization was performed using $\lambda=0.01$, $\mu$ = 1, and zero initialization.  The optimization was performed using in-house MATLAB software.

To demonstrate the importance of spatial constraints, we also estimated 2D spectra voxel-by-voxel using conventional 2D correlation spectroscopy techniques (i.e., without the spatial constraint). In addition, for comparison against 1D relaxometry, we simulated 1D $T_1$ relaxometry with the same 7 inversion times as in the high-dimensional case and  1D $T_2$ relaxometry with 32 echo times ($TE$s ranging from 10 ms to 320 ms) as in conventional approaches, with appropriate data averaging so that experiment durations were always matched.  Each 1D spectroscopic image was estimated using the same basic optimization formulation from Eq.~\eqref{eq:mat} (including spatial regularization to improve the estimation results), but modified so that we only had a 1D $T_1$ or $T_2$ spectrum at each voxel. 

\begin{figure}[tp]
	\centering
	{\includegraphics[width=0.7\linewidth]{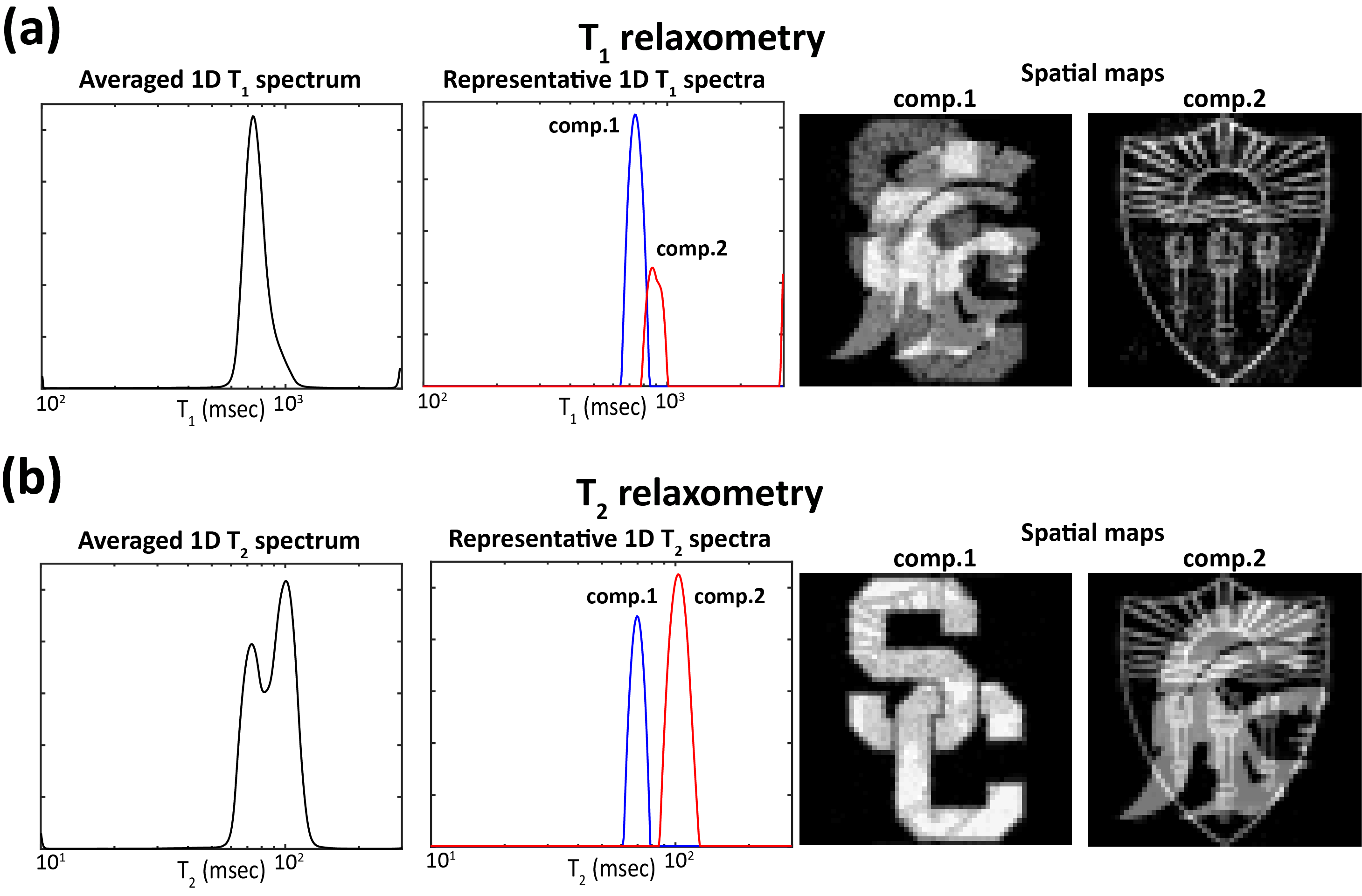}}
	
	\caption{Estimation results corresponding to simulated  (a) 1D $T_1$ relaxometry and (b) 1D $T_2$ relaxometry acquisitions.  Each figure shows (left) the  1D spectra obtained by spatially-averaging the 3D spectroscopic image, (middle) representative single-voxel spectra, and (right) spatial maps obtained by spectrally-integrating the spectral peak locations.} \label{fig:simulation_1D} 		
\end{figure}

Figures~\ref{fig:simulation} (c) and (d) show 2D correlation spectroscopic imaging results from our spatial-spectral approach as well as  conventional voxel-by-voxel 2D spectrum estimation.  As can been seen in Fig.~\ref{fig:simulation}(c), two strong peaks and one weak peak are discernible in the spatially-averaged 2D $T_1$-$T_2$ spectrum from our approach, and spatial maps obtained by spectral integration of these peaks are well matched to the ground truth spatial maps.  However, the reconstructed spectral peak widths were broader than the ground-truth peaks, as should be expected based on the use of finite sampling and the resolution limits imposed by the ill-posedness of multi-exponential signal estimation \cite{bertero1982}.   

In contrast, as can been seen in Fig.~\ref{fig:simulation}(d), we observe that the 2D correlation spectrum is not as well depicted when using voxel-by-voxel 2D spectrum estimation, with potentially a fourth peak emerging in between the peaks ascribed to components 1 and 2.  As can be seen,  the spatial maps for each component also exhibit cross-contamination, where the spatial details have bled from one component to another, suggesting a lack of adequate spectral resolution.

Another important observation is that our spatial-spectral estimation approach successfully resolves the fine spatial details of compartment 3, while voxel-by-voxel estimation was substantially less successful.  This occurred despite the fact that compartment 3 is not very spatially smooth, while spatial smoothness constraints are the only difference between the our approach and the conventional voxel-by-voxel method.  These results empirically demonstrate the benefits of the spatial-spectral approach to this inverse problem, and underscore the fact that the true spectroscopic image does not actually need to be very spatially smooth for these constraints to be useful.

For comparison, results from the 1D relaxometry simulations are shown in Fig.~\ref{fig:simulation_1D}.  As expected based on our previous analyses, both of the 1D relaxometry approaches fail to resolve three distinct spectral peaks, and were substantially less successful than the 2D approaches at recovering the spatial maps of the original three components.

\subsection{Pumpkin Experiment}
Real MRI data of a small pumpkin was acquired at every combination of 7 inversion times ($TI$ = 0, 100, 200, 400, 700, 1000 and 2000 ms) and 15 echo times ($TE$s ranging from 7.5ms to 217.5 ms in 15 ms increments) for a total of $P$ = 105 contrast encodings.  We used an IR-MSE sequence on a 3T MRI scanner (Achieva; Philips Healthcare, Best, The Netherlands).  For each $TI$, this sequence uses a train of spin-echoes to acquire data from all 15 different $TE$s in a single shot after the initial inversion recovery preparation. Acquisition used the following imaging parameters: 2 mm $\times$ 2 mm in-plane resolution, 4mm slice thickness, $TR$ = 5000 ms, a 32-channel receiver array coil, and SENSE parallel imaging with an acceleration factor of 2.  A high-resolution image of the same slice was additionally acquired with 0.5 mm $\times$ 0.5 mm in-plane resolution for reference.  For spectroscopic image estimation, we used the same dictionary matrix $\mathbf{K}$ and the same optimization parameters that were used in the numerical simulation.

It should be noted that $TI=0$ was not practical to implement, but it would theoretically produce the same sequence of magnitude images as a standard multi-echo spin-echo sequence without an inversion pulse.  As a result, we acquired data corresponding to $TI=0$ without using an inversion pulse and manually inverted the signal polarity.  This procedure assumes perfect inversion of the longitudinal magnetization.  However, the rest of data was acquired with a real inversion pulse, for which inversion efficiency may not be perfect due to various factors.  In addition, the scanner also used a different (and unknown) scaling factor when saving images with an inversion pulse than it did when saving images without.  As a result, the data corresponding to $TI$ = 0 had different scaling compared to the rest of the data. To correct for this unknown scale factor, we fit a single-compartment (monoexponential) inversion recovery curve to the data acquired with $TI>0$, and used this model to synthesize what the signal should have looked like at $TI=0$.  We compared this synthesized data against the measured data at $TI=0$ to generate a scale factor for each voxel.  A single global scale correction was then obtained by averaging these voxelwise scale factors, and applied to the measured data at $TI=0$.

To compare our multidimensional correlation spectroscopic imaging approach against 1D methods, we also performed both 1D $T_1$ relaxation and 1D $T_2$ relaxation spectrum estimation. In particular, 1D $T_1$ relaxation spectra were estimated for every voxel from the seven $TI$s at $TE$ = 7.5 ms, and 1D $T_2$ relaxation spectra were estimated for every voxel from the fifteen $TE$s.  Both 3D spectroscopic images (i.e., 1D spectra at every voxel) were estimated using the same estimation parameters described in the numerical simulation, including spatial regularization to improve the estimation results.

\begin{figure}[tp]
	\centering
	{\includegraphics[width=.7\linewidth]{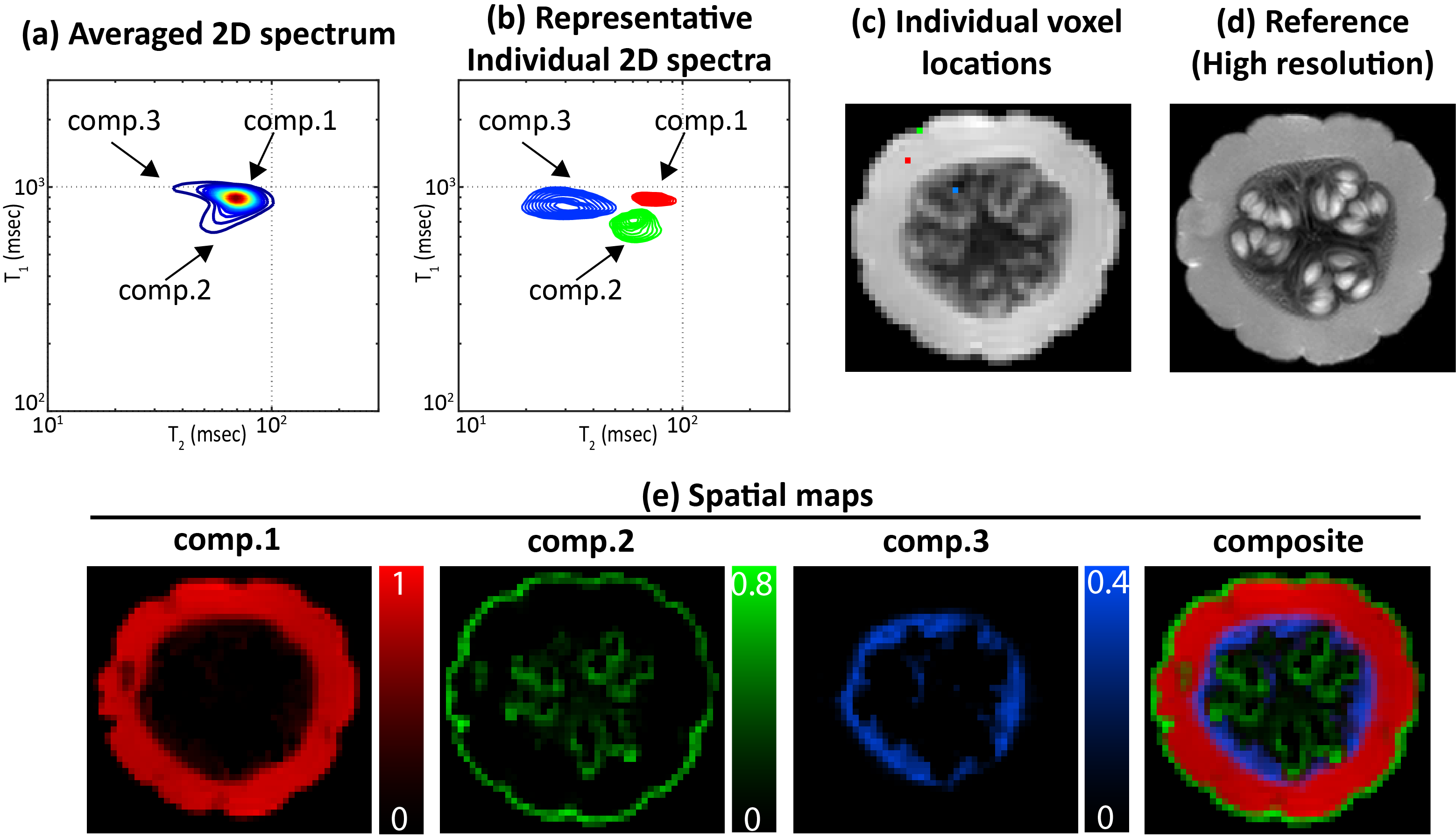}}
	\caption{Estimation results from the pumpkin experiment.  (a) The 2D spectrum obtained by spatially-averaging the estimated 4D spectroscopic image.  (b)  Representative individual spectra plotted from three different spatial locations.  In this plot, each of the spectral peaks has been numbered and color-coded (red: component 1, green: component 2 and blue: component 3).  The individual spatial locations corresponding to each of the peaks are depicted using the same color-coding scheme shown in (c) a representative image acquired at $TI$ = 0 ms and $TE$ = 7.5 ms.  (d) A high-resolution (0.5 mm $\times$ 0.5 mm) reference image.  (e)  Spatial maps obtained by spectrally-integrating the three spectral peaks.  Each of the map is color-coded based on the color-coding scheme described in (b), with the composite image  shown on the right.  }\label{fig:pumpkin_rrcsi}
\end{figure}

Figure~\ref{fig:pumpkin_rrcsi} shows the results from our multidimensional correlation spectroscopic imaging approach.  As shown in Fig.~\ref{fig:pumpkin_rrcsi}(a), we visually (subjectively) identified one strong peak and two weak peaks are resolved in the spatially-averaged 2D spectrum, and these three peaks are even more clearly distinguished when looking at the spectra from representative individual voxels as shown in Fig.~\ref{fig:pumpkin_rrcsi}(b).  By spectrally-integrating these three peaks, spatial maps were generated as shown in Fig.~\ref{fig:pumpkin_rrcsi}(e).  The three compartments that are observed in each of the spatial maps are consistent with high-resolution features from the reference image in Fig.~\ref{fig:pumpkin_rrcsi}(d), and while we do not claim  to be pumpkin experts, they appear to be consistent with our understanding of the anatomical structure of the pumpkin. 

\begin{figure}[tp]
	\centering
	{\includegraphics[width=.7\linewidth]{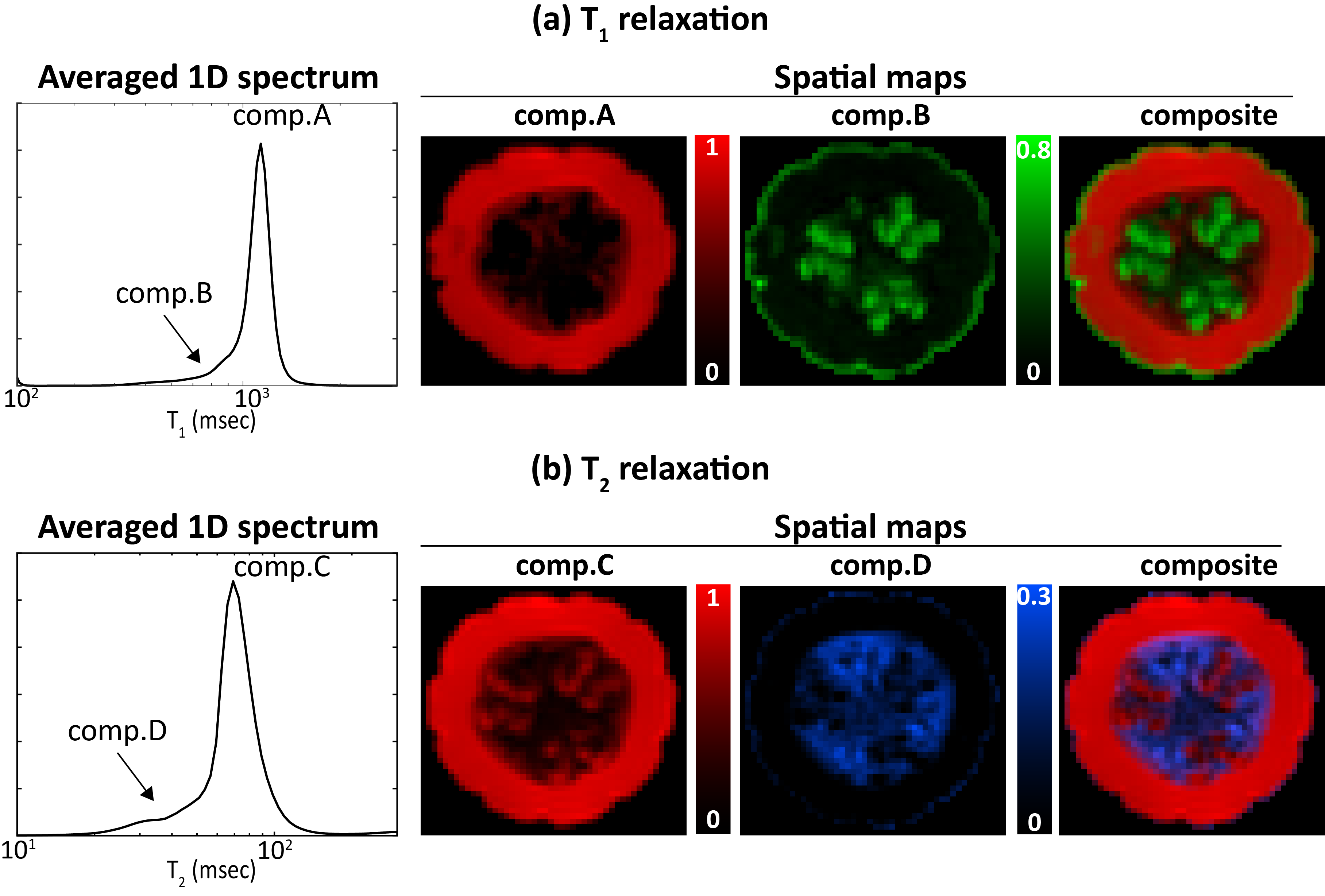}}
	\caption{Pumpkin experiment results from (a) conventional 1D T1 relaxometry and (b) conventional 1D T2 relaxometry. Each figure shows (left) the estimated spectra averaged across all voxels, and (right) spatial maps of the integrated spectral peaks.
	}\label{fig:pumpkin_1D}
\end{figure}

For comparison, Fig.~\ref{fig:pumpkin_1D} shows results from conventional 1D methods.  While the three spectral peaks were well-resolved in 2D, only two peaks are observed for both of the 1D methods.  Spatial maps of these peaks further illustrate that the conventional 1D methods do not resolve compartments successfully as our multidimensional correlation spectroscopic imaging approach.

\subsection{In vivo Human Brain Experiments}
We also acquired in vivo human brain data using the same imaging protocol and sequence parameters from the pumpkin experiment.  Axial slices with 4 mm thickness and coronal slices with 2 mm thickness were acquired from four healthy subjects (3 females and 1 male, and age: 30 $\sim$ 55 years).  We acquired one axial and one coronal slice from subject 1, two axial slices and two coronal slices from subject 2, and two axial slices from both subjects 3 and 4.  Contrast encoding used the same 105 (TE,TI) combinations as for the pumpkin experiment.  Each single-slice dataset was acquired within 20 minutes.  Just like for the pumpkin experiment, scale correction was performed for the data at $TI=0$, and then spectroscopic image reconstruction was performed using the same parameters described previously. 

To enable comparison against a 1D relaxometry method, we also used a multi-echo spin-echo sequence to acquire a $T_2$ relaxometry dataset from one subject with 32 $TE$s ranging from 10 ms to 320 ms in 10 ms increments, and otherwise using the same imaging parameters described previously.  This set of sequence parameters is typical for $T_2$-based myelin water imaging \cite{Mackay1994,Whittall1997}, and we chose to compare against this case because multicomponent $T_2$ relaxometry is more common in the literature than multicomponent $T_1$ relaxometry.   Estimation of the 3D spectroscopic image was performed using the same basic optimization formulation from Eq.~\eqref{eq:mat} (including spatial regularization), but the spectroscopic image and the dictionary matrix were modified for 1D $T_2$ relaxation.

\begin{figure}[tp]
	\centering
	{\includegraphics[width=.7\linewidth]{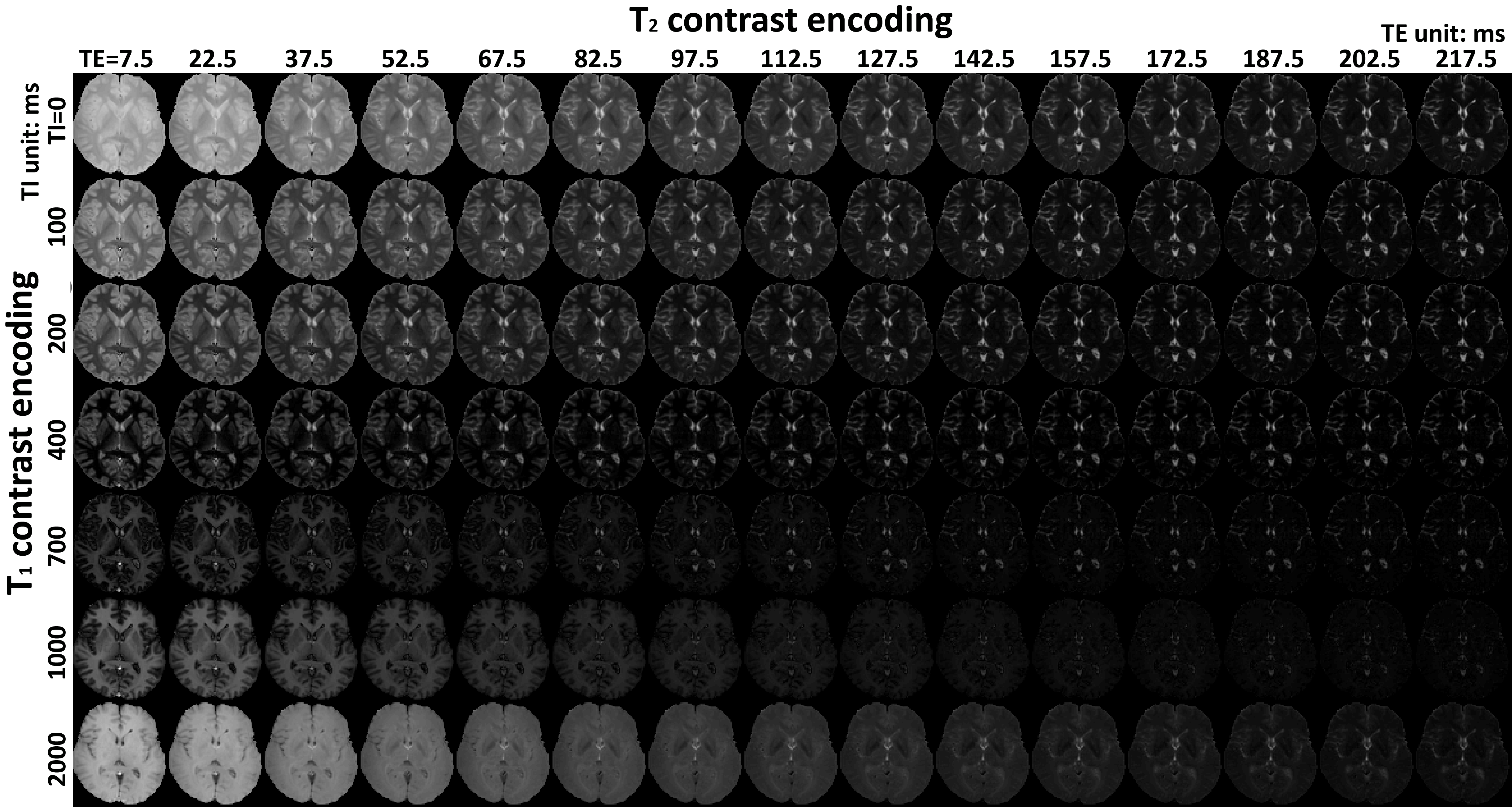}}
	\caption{A representative single-slice 4D dataset from an in vivo human brain (the axial slice from subject 1).}\label{fig:rrcsi_dataset_full}
\end{figure}

\begin{figure}[tp]
	\centering
	{\includegraphics[width=.7\linewidth]{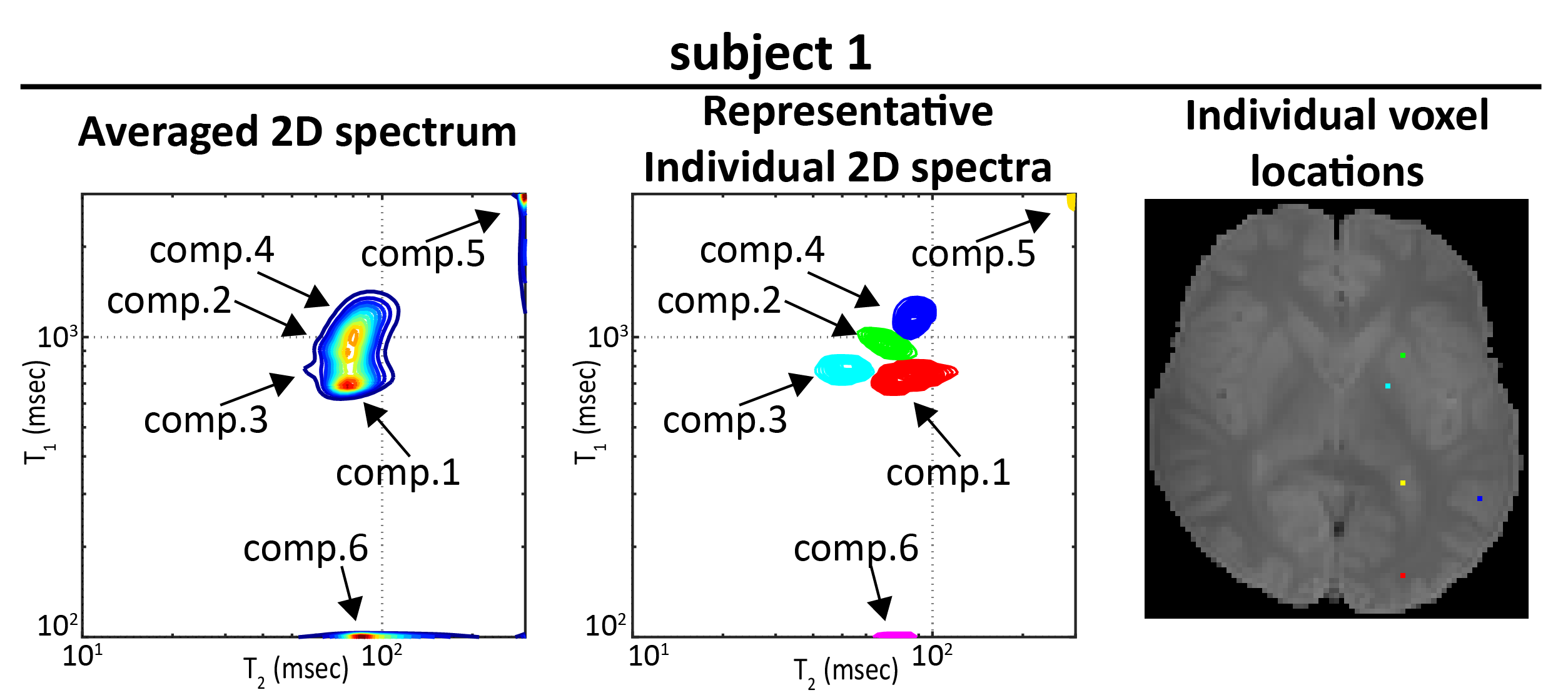}}
	
	\caption{2D $T_1$-$T_2$ correlation spectra estimated from an in vivo human brain (the axial slice from subject 1). (left) The 2D spectrum obtained by spatially-averaging the estimated 4D spectroscopic image.  (middle) Representative individual spectra plotted from different voxels.  Component 1 and component 6 are plotted from a white matter voxel, component 2 is plotted from a voxel in the putamen, component 3 is plotted from a voxel in the globus pallidus, component 4 is plotted from a voxel in gray matter, and component 5 is plotted from a voxel in the cerebral spinal fluid.  In this plot, each of the spectral peaks has been numbered and color-coded (red: component 1, green: component 2, cyan: component 3, blue: component 4, yellow: component 5, and magenta: component 6. This color coding scheme was also used to depict the individual voxel locations on (right) an anatomical reference image, although we do not mark the voxel for component 6 because it is the same as the voxel for component 1.} \label{fig:sub1_slice1} 		
\end{figure}

\begin{figure}[tp]
	\centering
	{\includegraphics[width=.7\linewidth]{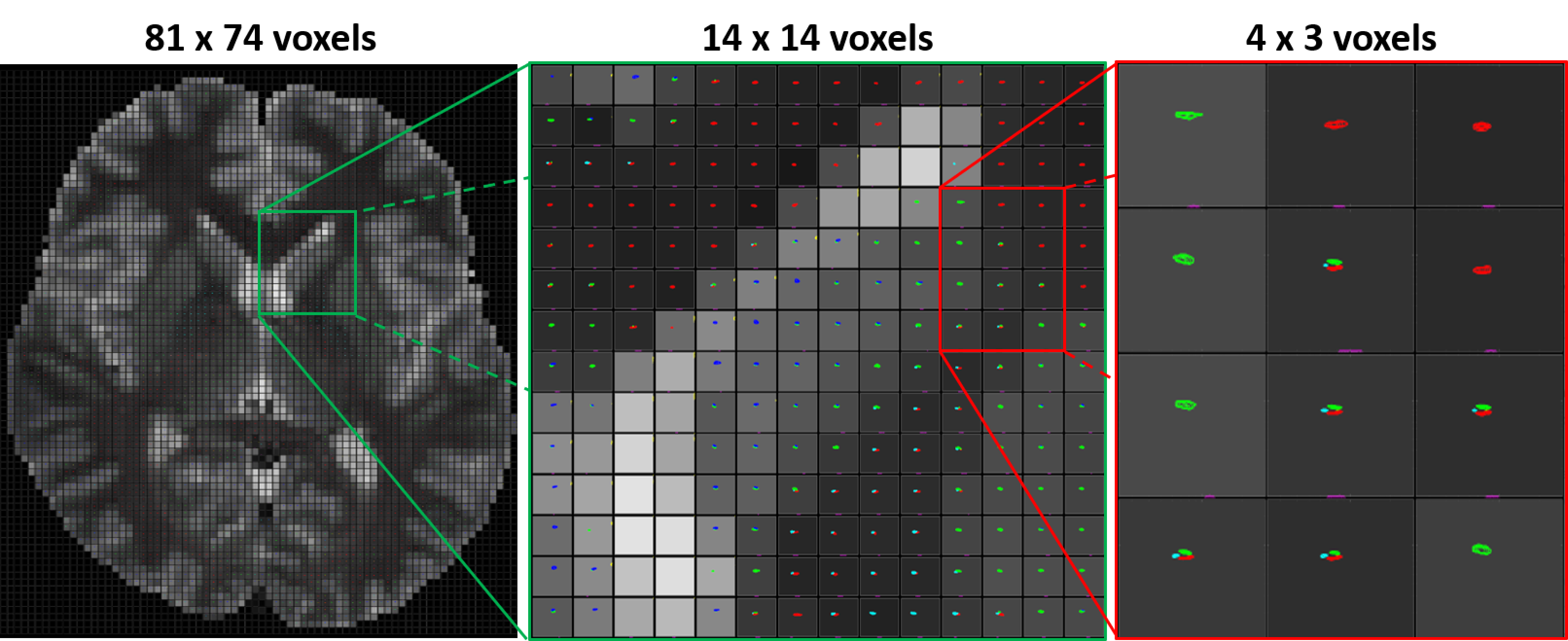}}
	
	\caption{Visualization of the estimated 4D spectroscopic image from an in vivo human brain (the axial slice from subject 1). Spatially-varying 2D spectra are shown from (left) the entire image (81 $\times$ 74 voxels), (middle) a subregion of the entire image corresponding to the green box (14 $\times$ 14 voxels), and (right) an even smaller subregion corresponding to the orange box (4 $\times$ 3 voxels). In each 2D spectrum, the horizontal axis corresponds to the $T_2$ dimension and the vertical axis corresponds to the $T_1$ dimension.  Each spectrum is color-coded based on the six spectral peaks and the color-coding scheme described in Fig.~\ref{fig:sub1_slice1}.} 
	\label{fig:spect_img}
\end{figure}

For illustration, a representative single-slice 4D dataset is shown in Fig.~\ref{fig:rrcsi_dataset_full} with corresponding 2D $T_1$-$T_2$ correlation spectra shown in Fig.~\ref{fig:sub1_slice1} and a visualization of the spectroscopic image shown in Fig.~\ref{fig:spect_img}.

As can be seen in Fig.~\ref{fig:sub1_slice1}, we visually (subjectively) identified six spectral peaks from the estimated 4D spectroscopic image. (While it may be difficult to distinguish the number of peaks that are present in the spatially-averaged spectrum,  the individual peaks are actually much easier to identify when looking at the spectra from individual voxels, as can be seen in both Figs.~\ref{fig:sub1_slice1} and \ref{fig:spect_img}.)   We believe that the capability to resolve six peaks is encouraging, since conventional 1D relaxometry methods generally only resolve two or three different compartments in the brain.  The ability of our multidimensional correlation spectroscopic imaging approach  to resolve substantially more spectral peaks is consistent with our expectations about the superiority of high-dimensional encoding and spatial-spectral estimation relative to lower-dimensional approaches.

As can be seen in Fig.~\ref{fig:spect_img}, we also frequently observe multiple peaks coexisting within a single voxel, and the peaks each have their own distinct spatial distributions.  If we ascribe the different spectral peaks to different microstructural tissue compartments, then we can interpret this 4D spectroscopic image as demonstrating the ability to resolve partial voluming and to spatially map the spatial variations of each compartment.

\begin{figure}[tp]
	\centering
	{\includegraphics[width=1\linewidth]{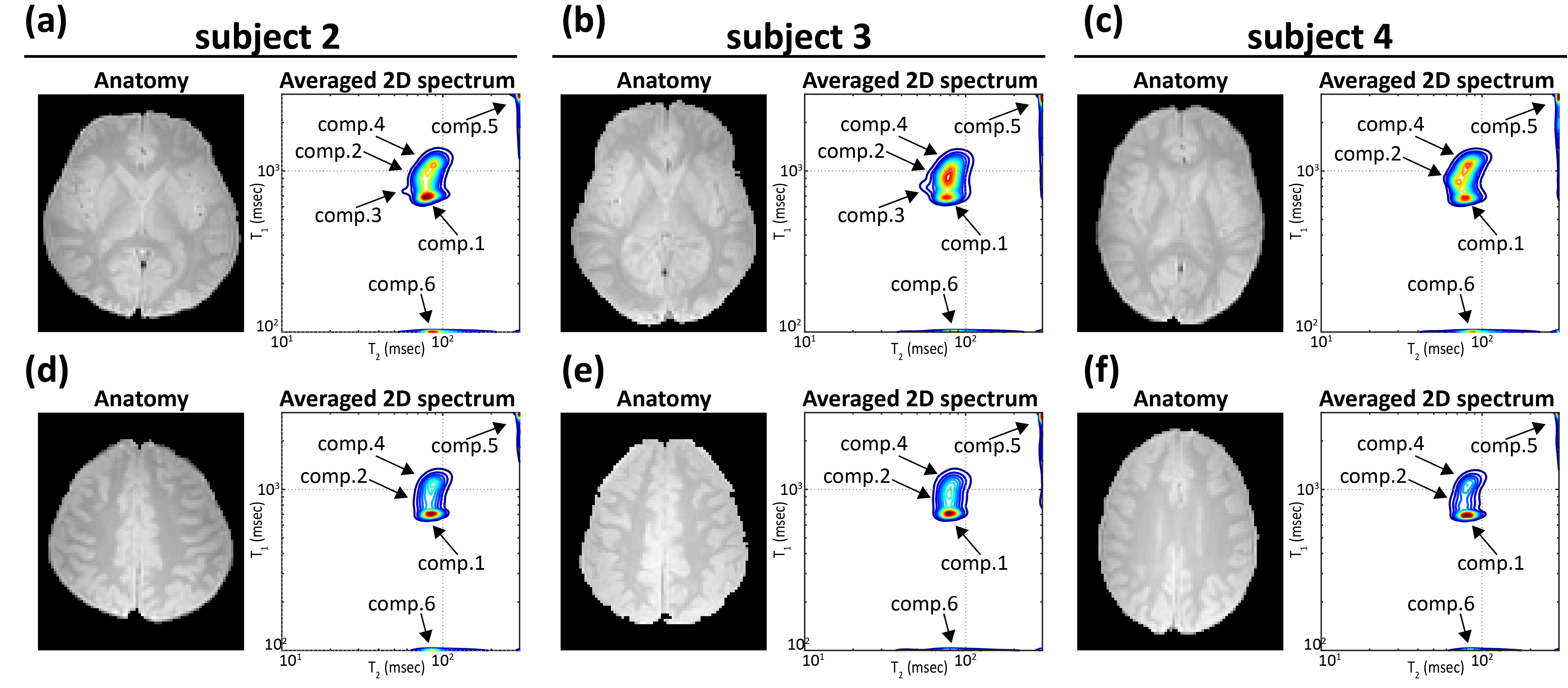}}
	
	\caption{Reference images (corresponding to  $TI$ = 0 and $TE$ = 7.5 ms) and spatially-averaged 2D $T_1$-$T_2$ spectra from different axial slices of different subjects.} \label{fig:spectra_axial} 
\end{figure}

2D spectra obtained by spatially averaging the 4D $T_1$-$T_2$ spectroscopic image are shown for different slices and subjects in Fig.~\ref{fig:spectra_axial} (axial slices) and Supporting Information~\ref{fig:spectra_coronal} (coronal slices). The number and spectral locations of the observed spectral peaks are largely the same as observed in the spectra shown in Fig.~\ref{fig:sub1_slice1}, demonstrating that our multidimensional correlation spectroscopic imaging approach appears to yield robust and consistent results across a range of different subjects and slice orientations.

\begin{figure}[tp]
	\centering
	{\includegraphics[width=.7\linewidth]{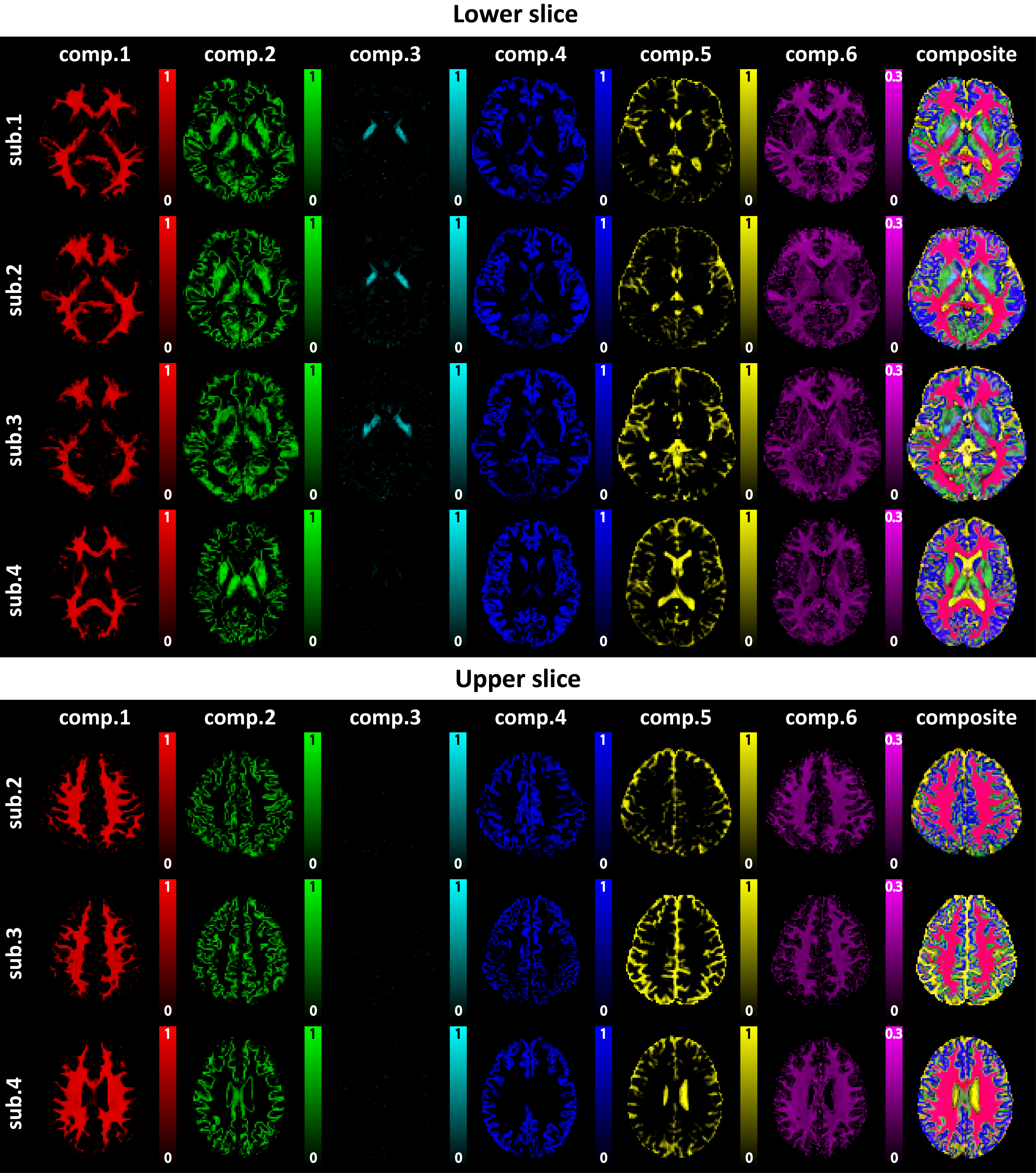}}
	
	\caption{Spatial maps obtained from our multidimensional correlation spectroscopic imaging approach by spectrally-integrating the six spectral peaks from axial slices of different subjects. Each map is color-coded based on the scheme from Fig.~\ref{fig:sub1_slice1}, and the composite image is also shown on the right.} \label{fig:maps_axial_slice1}
\end{figure}

Spatial maps obtained by spectrally integrating the six previously-identified spectral peaks are shown in Fig.~\ref{fig:maps_axial_slice1} (axial slices) and Supporting Information.~\ref{fig:maps_coronal} (coronal slices).  We observe that these maps are also largely consistent with one another.

Importantly, the spatial maps also appear to qualitatively match well with known brain anatomy: component 1 seems to correspond to a white matter (WM) compartment; component 2 seems to correspond to GM structures with relatively high myelin content, including subcortical GM, putamen, thalamus and brainstem nuclei (as seen on the wall of the fourth ventricle in subject 1 in Fig.~\ref{fig:maps_coronal}) as well as cortical GM; component 3 seems to correspond to brain structures with high iron content including the globus pallidus, subthalamic nucleus and substantia nigra; component 4 is similar to component 2, but seems to represent the GM content absent any myelin-content and notably does not include the subcortical GM; component 5 seems to correspond to cerebrospinal fluid (CSF); and component 6 resembles the myelin water compartment that has been observed in previous myelin water imaging experiments \cite{Mackay1994,Whittall1997,Labadie1994,Oh2013}. 

It should be noted that the component 3 is not observed in every slice, which we believe is reasonable because the third compartment seems localized to gray matter structures like the globus pallidus, and these structures are not present in all of the slices we acquired data from.

It should also be noted that the relaxation parameter values we estimated for component 5 (which seems to correspond to CSF) and component 6 (which resembles  myelin water) do not match the parameter values for these tissue types reported in previous literature \cite{mackay2006,Does1998, Does2002, Does2018}.  This is somewhat expected for a variety of reasons (including simplistic modeling assumptions as will be discussed in the next section), but is especially expected because the range of contrast encoding parameters we used may be insufficient to accurately estimate very quickly-relaxing tissues like myelin water or very slowly-relaxing tissues like CSF.  Nevertheless, while the specific relaxation parameter values we've estimated are unlikely to be accurate, we are encouraged by the fact that it appears that our multidimensional correlation spectroscopic imaging approach may still be successfully resolving the spatial maps of these components, and that the resulting maps are still consistent from slice to slice and subject to subject.

\begin{figure}[tp]
	\centering
	{\includegraphics[width=.6\linewidth]{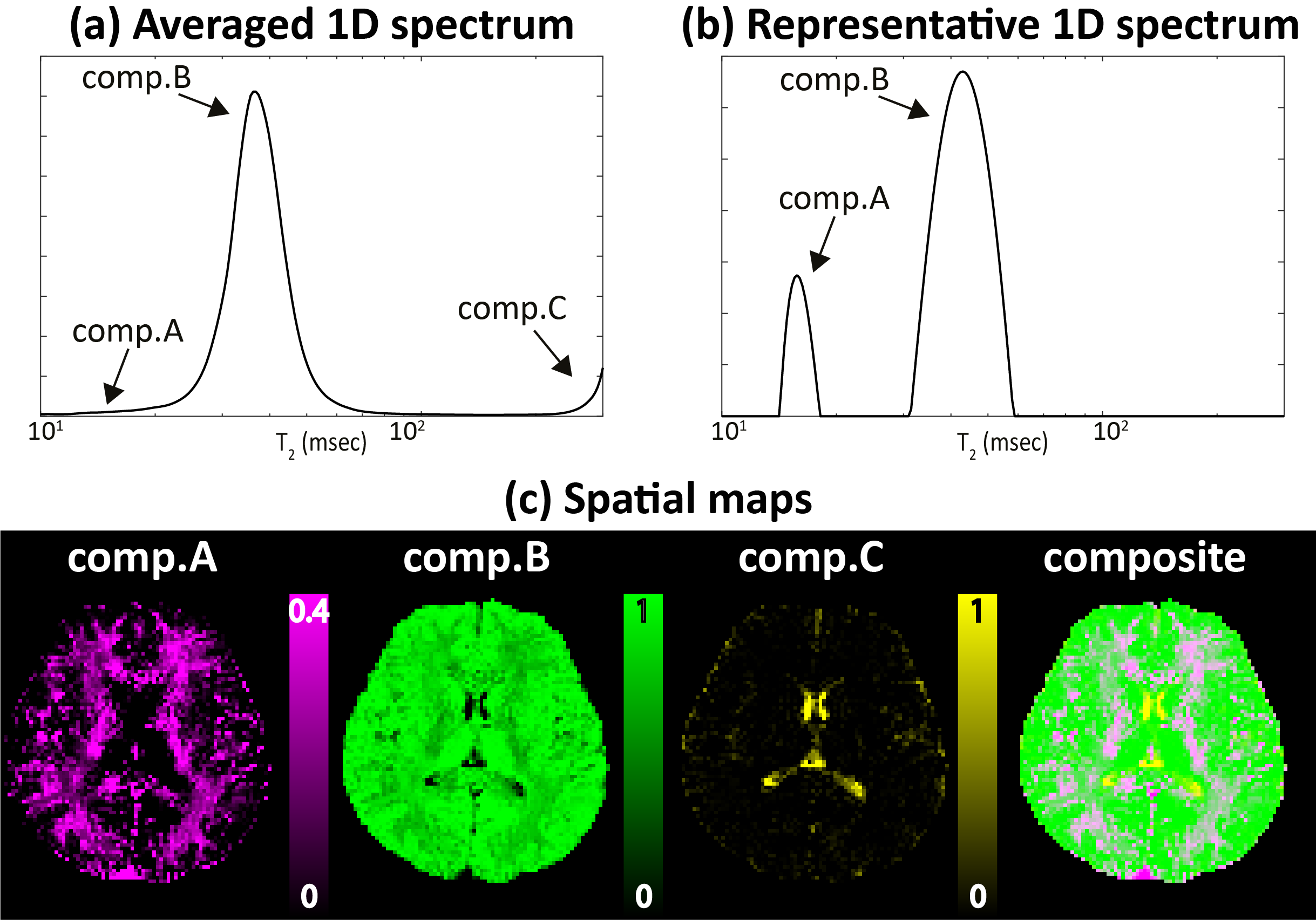}}	
	
	\caption{Results from conventional 1D $T_2$ relaxation. (a) The 1D spectrum obtained by spatially-averaging the 3D spectroscopic image.  (b) Representative 1D spectrum  from a voxel in white matter. (c) Spatial maps obtained by spectrally-integrating the three spectral peak locations. Each map is color-coded (magenta: comp.A, green: comp.B, and yellow: comp.C), and the composite image is also shown on the right.} \label{fig:conventional_t2}
\end{figure}

For comparison, Fig.~\ref{fig:conventional_t2} shows the results obtained from the 1D $T_2$ relaxometry experiment. As can been seen (and consistent with previous literature \cite{Mackay1994,Whittall1997}), only three spectral peaks are resolved, which is substantially fewer than the number of peaks resolved by our multidimensional correlation spectroscopic imaging approach.  While the spatial maps corresponding to these three peaks all appear to be anatomically reasonable, we believe that the interpretation of these maps is less straightforward than the interpretation of the spatial maps from our approach.  In particular, we believe that the 1D relaxometry results are still substantially confounded by partial volume contributions, which appear to be more successfully resolved by 2D $T_1$-$T_2$ correlation spectroscopic imaging.

\section{Discussion}\label{discussion}

This work described our approach to multidimensional correlation spectroscopic imaging and demonstrated empirically that this approach can have better compartmental resolving power than lower-dimensional approaches.  However, using higher-dimensional contrast encoding is associated with practical increases in data acquisition time.  While our human brain experiments were reasonably fast, a 20 minute acquisition may still be too long for routine practical use of this technique.  However,  there are still plenty of opportunities to make the scan faster.  For example, while our acquisition used a relatively large number (i.e., 105) of different contrast encodings, it should be noted that this number of samples was not optimized, and was chosen based on the maximum number of contrasts we could fit within a 20 minute acquisition time.  We have recently presented preliminary work that uses CRLB theory to optimize experimental protocols for both  2D $T_1$-$T_2$ \cite{Kim2017assilomar} and 2D $D$-$T_2$ \cite{Kim2017ismrm_a} correlation spectroscopic imaging. This preliminary work has demonstrated that we can obtain similar-quality results with substantially less than 105 encodings, which may be leveraged to enable substantial improvements in data acquisition time.   In addition, there have been other recent constrained reconstruction approaches that have been proposed to reduce the contrast-encoding requirements of high-dimensional relaxation spectroscopy \cite{Bai2015, Benjamini2016, Benjamini2017}, as well as constrained reconstruction approaches that have been proposed to reduce the k-space sampling and averaging requirements of multi-contrast imaging \cite{haldar2008a,gong2015,zhao2015,bilgic2018}.  Simultaneous-multislice imaging \cite{barth2016} could also be used to increase the volume coverage of an acquisition without increasing the acquisition time.  In addition, while we relied on an IR-MSE sequence for the results shown in this paper, our approach can also be used with  other pulse sequences that may have higher efficiency such as MR fingerprinting \cite{Ma2013}, inversion recovery balanced steady-state free-precession \cite{Schmitt2004, Pfister2018}, or a recent advanced high-dimensional contrast encoding technique \cite{Hutter2018}.  Multicomponent MR fingerprinting methods have also recently been described in the literature \cite{mcgivney2018,Tang2018} that have similar objectives to 2D $T_1$-$T_2$ correlation spectroscopic imaging, although it does not appear that these approaches were able to resolve as many tissue compartments as our approach does.  Recently, we have presented preliminary results of 2D $T_1$-$T_2$ correlation spectroscopic imaging with an MR fingerprinting acquisition with very promising results \cite{Kim2019ismrm}.  Any of these kinds of approaches could potentially be synergistically combined to make multidimensional correlation spectroscopic imaging experiments even faster.

Although our experimental results appear to demonstrate  successful decomposition of sub-voxel compartments from in vivo human subjects, some of the estimated relaxation parameters we estimated do not match closely with previous literature values.  As explained previously, some of these discrepancies should be expected because our range of TEs and TIs may not be sufficient to accurately estimate very long or very short relaxation parameters.  However, it is also known that quantification of relaxation parameters can be significantly affected by a variety of factors including experimental conditions, signal modeling, and optimization parameters, leading to mismatches even between different standard approaches \cite{Stikov2014}.  Recognizing these issues, we have focused on qualitative evaluation in real data scenarios rather than quantitative validation because of the lack of a gold standard.  Nevertheless, we were able to show promising results in terms of reproducibility and consistency throughout the experiment of multiple subjects.

Although we were able to get consistently-similar and qualitatively-reasonable correlation spectroscopic imaging results, the ill-posedness of the inverse problem still means that we do not expect that our reconstructions will necessarily be the unique spectroscopic images that are consistent with the measured data.  As a result, it would likely be fruitful in the future to do further exploration of the uncertainty of our solutions to the inverse problem, e.g., as might be achieved using optimization tools\cite{parker2005} or Monte Carlo methods\cite{prange2009,prange2010}.)

Similarly, it is also worth noting that our application examples with the IR-MSE sequence used a relatively simple acquisition and a substantially simplified physics model that does not account for the effects of water exchange, magnetization transfer, B1 inhomogeneity, B0 inhomogeneity, slice profile effects, etc.  Thus, while our current implementation appears to successfully enable separation of tissue compartments that are not easily resolved by other methods, we believe that histological validation and improving the quantitative accuracy of this kind of approach are important future objectives.   (Note that it has been suggested that the standard nonnegativity constraint may be problematic for IR-MSE sequences in the presence of significant exchange  \cite{Dortch2009}, such that extending our approach to account for exchange may require additional modifications to our constrained estimation framework.)   More accurate modeling  may also enable the use of more advanced pulse sequences that may be more efficient than IR-MSE sequences but are more sensitive to nonideal acquisition physics.  
  
Lastly, it should be noted that our estimation formulation from Eq.~\eqref{eq:mat} uses a least-squares penalty to enforce consistency.  The use of least-squares is appropriate under a Gaussian noise model, although the magnitude images we used in this work are more properly modeled with a Rician distribution.  While the differences between Gaussian noise and Rician noise are unlikely to matter very much in high-SNR regimes, it may be valuable to investigate the use of more accurate statistical noise modeling approaches \cite{varadarajan2015}  in future work.

\section{Conclusion}\label{conclusion}
This work presented an overview of our approach to multidimensional correlation spectroscopic imaging of exponential decays, and demonstrated a range of theoretical, simulation, and experimental results to illustrate the benefits of this approach.  This approach combines high-dimensional contrast encoding with high-dimensional spatial-spectral image reconstruction to reduce the ill-posedness associated with separating multiple sub-voxel tissue compartments, enabling good results without requiring an extensive amount of pristine data.  Our results demonstrate the strong potential of the method using both numerical simulations and real MRI data, including what we believe are the first in vivo human experiments of this kind.  While we demonstrated 2D $T_1$-$T_2$ experiments in this work and 2D $D$-$T_2$ experiments in our previous work\cite{Kim2017}, we are excited by the potential of this type of approach to be used with a wider range of MR contrast parameters and/or even higher-dimensional contrast encoding.

\section*{Acknowledgments}
This work was supported in part by the USC Alfred E. Mann Institute and research grants NSF CCF-1350563, NIH R21 EB022951, NIH R01 MH116173, NIH R01 NS074980, NIH R01 NS089212, and NIH R21 EB024701.  Computation for some of the work described in this paper was supported by the University of Southern California's Center for High-Performance Computing (http://hpcc.usc.edu/).  The authors thank Dr. Quin Lu from Philips Healthcare for clinical science support, and thank Dr. Jongho Lee for helpful comments on the manuscript.  

\section*{Supporting information}

The following supporting information is available as part of the online article:

\noindent
\textbf{Figure S1.}
{ Reference images (corresponding to  $TI$ = 0 and $TE$ = 7.5 ms) and spatially-averaged $T_1$-$T_2$ spectra obtained from our multidimensional correlation spectroscopic imaging approach from different coronal slices of different subjects.}

\noindent
\textbf{Figure S2.}
{Spatial maps obtained from our multidimensional correlation spectroscopic imaging approach by spectrally-integrating the six spectral peaks from coronal slices  of different subjects. Each map is color-coded based on the scheme from Fig.~\ref{fig:sub1_slice1}, and the composite image is also shown on the right.}

\appendix

\section{ADMM Based Optimization Algorithm} \label{app:appendixA}

In this section, we describe an ADMM based optimization algorithm to solve Eq.~\eqref{eq:optimization}.  Compared to the algorithm described in our previous work  \cite{Kim2016isrmrm,Kim2017,Kim2017spie,Kim2018ismrm, Kim2018isbi} (which used vector-representations of the data and the spectroscopic image within each computation step), the description below instead relies on matrix representations of the variables in Eq.~\eqref{eq:optimization} as in Eq.~\eqref{eq:mat}.  This is not just a notational difference, because computing each step using the matrix representation in Eq.~\eqref{eq:mat} removes the need for for-loops and reduces computational complexity.

The algorithm proceeds as follows:

\begin{itemize}
	\item Set iteration number $j=0$, and initialize $\hat{\mathbf{F}}^{(0)}\in\mathbb{R}^{Q\times N}$, $\hat{\mathbf{X}}^{(0)}\in\mathbb{R}^{Q\times N}$, $\hat{\mathbf{Y}}^{(0)}\in\mathbb{R}^{Q\times N}$, $\hat{\mathbf{Z}}^{(0)}\in\mathbb{R}^{Q\times N}$, $\mathbf{G}^{(0)}\in\mathbb{R}^{Q\times N}$, $\mathbf{H}^{(0)}\in\mathbb{R}^{Q\times N}$ and $\mathbf{R}^{(0)}\in\mathbb{R}^{Q\times N}$ to arbitrary values. The variables $\mathbf{X}$, $\mathbf{Y}$, and $\mathbf{Z}$ are used for variable splitting, while the variables $\mathbf{G}$, $\mathbf{H}$, and $\mathbf{R}$ correspond to Lagrange multipliers.    Also choose an augmented Lagrangian parameter value $\mu >0$ (this choice influences the convergence speed of ADMM but not the solution).
	\item At iteration $(j+1)$:
	\begin{enumerate}
		\item For each $i=1,\ldots,N$, update the estimate of $\hat{\mathbf{f}}_i$ (i.e., the columns of $\hat{\mathbf{F}}$) according to
		\begin{equation}
		\hat{\mathbf{f}}_i^{(j+1)} =\left\{\begin{array}{ll} \frac{1}{3}\left(\hat{\mathbf{x}}_i^{(j)}+{\mathbf{g}}_i^{(j)} + \hat{\mathbf{y}}_i^{(j)} + {\mathbf{h}}_i^{(j)} + \hat{\mathbf{z}}_i^{(j)} + {\mathbf{r}}_i^{(j)} \right),\\ \hspace{130pt} \text{ if } t_i = 1 \\ \frac{1}{2}\left(\hat{\mathbf{y}}_i^{(j)} + {\mathbf{h}}_i^{(j)} + \hat{\mathbf{z}}_i^{(j)} + {\mathbf{r}}_i^{(j)} \right), \\
		\hspace{130pt} \text{ if } t_i = 0,\end{array} \right.
		\end{equation}
		where $\hat{\mathbf{x}}_i^{(j)}$, ${\mathbf{g}}_i^{(j)}$, $\hat{\mathbf{y}}_i^{(j)}$, ${\mathbf{h}}_i^{(j)}$, $\hat{\mathbf{z}}_i^{(j)}$ and ${\mathbf{r}}_i^{(j)}$ are respectively the $i$th columns of $\hat{\mathbf{X}}^{(j)}$, ${\mathbf{G}}^{(j)}$, $\hat{\mathbf{Y}}^{(j)}$, ${\mathbf{H}}^{(j)}$, $\hat{\mathbf{Z}}^{(j)}$, and ${\mathbf{R}}^{(j)}$.
		\item Set
		\begin{equation} 
		\hat{\mathbf{X}}^{(j+1)} = \left(\mathbf{K}^H\mathbf{K}+\mu\mathbf{I}\right)^{-1}\left(\mathbf{K}^H\mathbf{M} +\mu\left(\hat{\mathbf{F}}^{(j+1)}\mathbf{T}-\mathbf{G}^{(j)}\right)\right),
		\end{equation}
		where  $\mathbf{I}$ denotes the identity matrix.  		Note that the matrix $\left(\mathbf{K}^H\mathbf{K}+\mu\mathbf{I}\right)$ is small and its inverse can easily be precomputed and stored for use in every iteration.
		\item Set $\hat{\mathbf{Y}}^{(j+1)} = \hat{\mathbf{F}}^{(j+1)} - {\mathbf{H}}^{(j)}$, but replacing any negative values with zero.						
		\item Set
		\begin{equation}
		\hat{\mathbf{Z}}^{(j+1)} = \mu\left(\hat{\mathbf{F}}^{(j+1)}-\mathbf{R}^{(j)}\right) \left(\mu\mathbf{I} + \lambda \mathbf{C}^H\mathbf{C}\right)^{-1}.
		\end{equation}
		Note that if we assume periodic boundary conditions for the spatial smoothness regularization, then the matrices $\mathbf{I}$ and $\mathbf{C}^H\mathbf{C}$ can both be diagonalized by the discrete Fourier transform, and we can use standard Fourier methods to quickly and analytically compute the desired matrix inversion result \cite{afonso2011}.
		\item Set
		\begin{equation}
		\begin{split}
		\mathbf{G}^{(j+1)} &= \mathbf{G}^{(j)} - ( \hat{\mathbf{F}}^{(j+1)}\mathbf{T} - \hat{\mathbf{X}}^{(j+1)}), \\ \mathbf{H}^{(j+1)} &= \mathbf{H}^{(j)} - ( \hat{\mathbf{F}}^{(j+1)} - \hat{\mathbf{Y}}^{(j+1)}),\\ \text{and } \mathbf{R}^{(j+1)} &= \mathbf{R}^{(j)} - ( \hat{\mathbf{F}}^{(j+1)} - \hat{\mathbf{Z}}^{(j+1)}).
		\end{split}
		\end{equation}
		\item Set  $j = j+1$.
	\end{enumerate}
	\item Iterate steps 1-6 until convergence.
\end{itemize}

\bibliography{./bibliography}

\clearpage
\setcounter{figure}{0} 
\renewcommand{\thefigure}{S\arabic{figure}}

\begin{figure}[t]
	\centering
	{\includegraphics[width=1\linewidth]{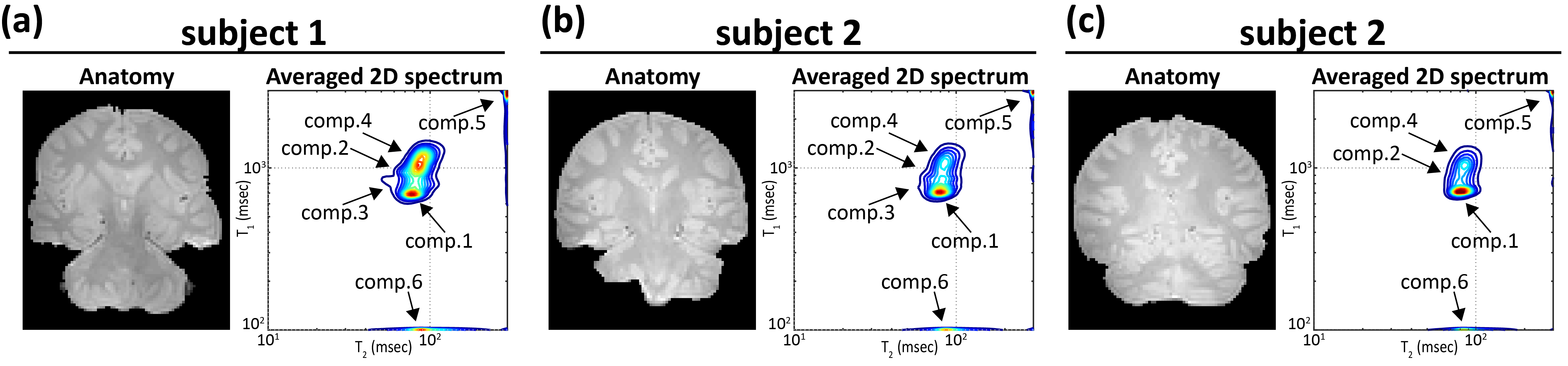}}
	
	\caption{ Reference images (corresponding to  $TI$ = 0 and $TE$ = 7.5 ms) and spatially-averaged $T_1$-$T_2$ spectra obtained from our multidimensional correlation spectroscopic imaging approach from different coronal slices of different subjects.} \label{fig:spectra_coronal}
\end{figure}
\clearpage

\begin{figure}[t]
	\centering
	{\includegraphics[width=1\linewidth]{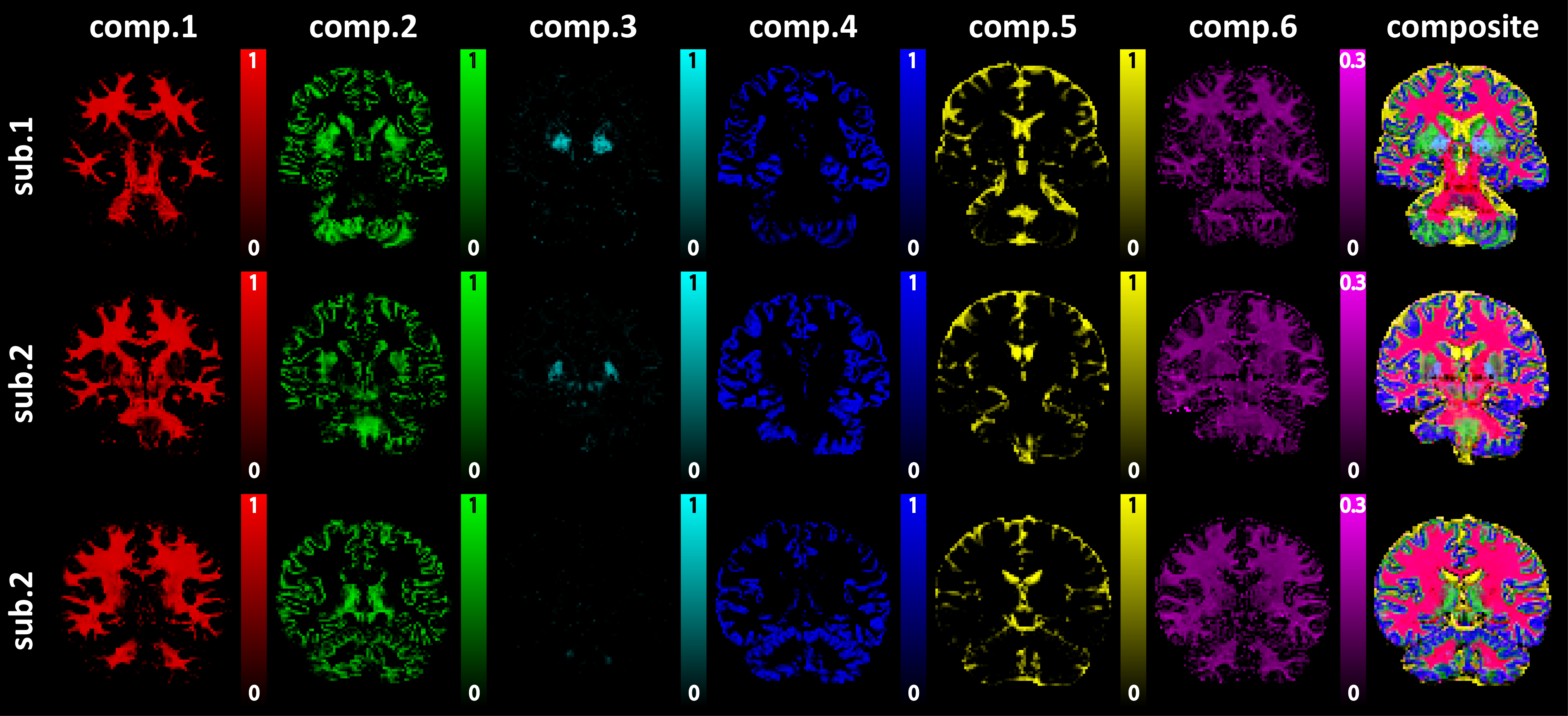}}
	
	\caption{Spatial maps obtained from our multidimensional correlation spectroscopic imaging approach by spectrally-integrating the six spectral peaks from coronal slices  of different subjects. Each map is color-coded based on the scheme from Fig.~\ref{fig:sub1_slice1}, and the composite image is also shown on the right.} \label{fig:maps_coronal}
\end{figure}

\end{document}